\documentclass[aps,prd,reprint,nofootinbib,amsmath,amssymb,superscriptaddress,longbibliography]{revtex4-1}

\usepackage{graphicx}   
\usepackage[
colorlinks=true,        
citecolor=blue,         
linkcolor=blue,         
urlcolor=blue ,          
]{hyperref}  
\usepackage{tabulary}
\usepackage{color}      
\usepackage{orcidlink}
\usepackage{multirow}
\usepackage{multirow}

\newcommand{\nc}{\newcommand*} 
\nc{\be}{\begin{equation}}
\nc{\ee}{\end{equation}}
\def\({\left(}
\def\){\right)}
\def\[{\left[}
\def\]{\right]}
\nc{\Eq}[1]{Eq.~\eqref{#1}}     
\nc{\Fig}[1]{Fig.~\ref{#1}}     
\nc{\Table}[1]{Table~\ref{#1}}  
\nc{\Sec}[1]{Sec.~\ref{#1}}     
\nc{\red}[1]{\textcolor{red}{#1}}

\newcommand{\aj}{Astron. J.} 
\newcommand{\apjl}{Astrophys. J. Lett.} 
\newcommand{\apjs}{Astrophys. J. Suppl. Ser.} 
\newcommand{\apss}{Astrophys. Space Sci.} 
\newcommand{\aap}{Astron. Astrophys.} 
\newcommand{\jcap}{J. Cosmol. Astropart. Phys.} 
\newcommand{\mnras}{Mon. Not. R. Astron. Soc.} 

\begin{document}

\title{Reconstructing the evolution of deceleration parameter with the Lagrange interpolation method}

\author{Kaituo Zhang\orcidlink{0000-0002-8653-3363}}
\affiliation{Department of Physics, Anhui Normal University, Wuhu, Anhui 241000, China}

\author{Bing Xu\orcidlink{0000-0002-9394-0426}}
\email{xub@ahstu.edu.cn}
\affiliation{School of Electrical and Electronic Engineering, Anhui Science and Technology University, Bengbu, Anhui 233030, China}

\author{Hongwei Yu\orcidlink{0000-0002-3303-9724}}
\email{hwyu@hunnu.edu.cn}
\affiliation{Department of Physics, Hunan Research Center of the Basic Discipline for Quantum Effects and Quantum Technologies, and Key Laboratory of Low-Dimensional Quantum Structures and Quantum Control of Ministry of Education,
Hunan Normal University, Changsha, Hunan 410081, China}

\author{Puxun Wu\orcidlink{0000-0002-9188-7393}}
\email{pxwu@hunnu.edu.cn}
\affiliation{Department of Physics, Hunan Research Center of the Basic Discipline for Quantum Effects and Quantum Technologies, and Key Laboratory of Low-Dimensional Quantum Structures and Quantum Control of Ministry of Education,
Hunan Normal University, Changsha, Hunan 410081, China}

\begin{abstract}
In this paper, we reconstruct the evolution of the deceleration parameter using cubic and quartic Lagrange interpolation methods, based on the latest observational data. This data includes baryon acoustic oscillation (BAO) measurements from Dark Energy Spectroscopic Instrument (DESI), cosmic microwave background (CMB) distance priors from Planck 2018, and three datasets of type Ia supernovae (SNe Ia): DES-Dovekie, Pantheon+, and Union3. To mitigate the impact of the nonuniform distribution of data in redshift, we carry out the reconstruction not only in the standard redshift $z$ but also in the $y$-redshift $y\equiv z/(1+z)$ and the log-redshift $\zeta\equiv\ln(1+z)$. The cubic and quartic reconstructions of $q(z)$ from the DESI BAO+CMB+SNe Ia datasets indicate a clear deviation from the cosmological constant plus cold dark matter ($\Lambda$CDM) model at low redshift ($z<0.3$), supporting a weaker present-day cosmic acceleration than the predictions of the $\Lambda$CDM model. The reconstructed $q(z)$ exhibits an overall oscillatory pattern around the $\Lambda$CDM model throughout its evolution. Only the cubic-$y$ and cubic-$\zeta$ reconstructions with the DESI BAO+CMB+Union3 data provide evidence that the cosmic acceleration appears to have already reached its maximum value and is beginning to slow down. However, when incorporating the measurement of the Hubble constant $H_0$ from the Supernovae and $H_0$ for the Equation of State (SH0ES) Collaboration, the evolution of $q(z)$ from cubic reconstructions shifts toward the $\Lambda$CDM prediction. Regardless of the presence or absence of the $H_0$ prior, all reconstructions imply that the deceleration-acceleration transition occurred earlier than predicted by the $\Lambda$CDM model. Nevertheless, the Bayesian evidence still favors $\Lambda$CDM, while the frequentist analysis indicates a $>2\sigma$ preference for the Lagrange-interpolation reconstructions, hinting at the existence of dynamical dark energy.

\end{abstract}

\maketitle


\section{Introduction}\label{s1}

The late-time accelerated expansion of the Universe has been strongly confirmed by various cosmological observations, including type Ia supernovae (SNe Ia)~\cite{Riess1998AJ,Perlmutter1999ApJ}, baryon acoustic oscillations (BAO)~\cite{Eisenstein2005ApJ,Percival2007MNRAS}, measurements of the Hubble parameter~\cite{Farooq2017ApJ,Yu2018ApJ}, and cosmic microwave background (CMB)~\cite{Komatsu2011ApJS,PlanckCollaboration2014A&A,PlanckCollaboration2020A&A}.  Before this phase of accelerating expansion, the Universe experienced a decelerating expansion era dominated by nonrelativistic matter, including cold dark matter and baryonic matter~\cite{Riess1998AJ,Perlmutter1999ApJ,Riess2004ApJ,Komatsu2011ApJS,PlanckCollaboration2014A&A,PlanckCollaboration2020A&A}. To effectively capture the characteristics of cosmic expansion, we need to study the evolution of the  deceleration parameter, defined as $q(z) = -\ddot{a}/(aH^2)$. This parameter encapsulates the dynamical effects of  matter and dark energy over time and thus provides a direct kinematic description of the cosmic expansion history. Here $a$ is the cosmic scale factor, $H=\dot{a}/a$ is the Hubble parameter and an overdot denotes a derivative with respect to the time $t$. Since the present-day value of $q(z)$, $q_0 = q(z=0)$, directly characterizes the current expansion status of the Universe, $q_0$, together with the Hubble constant $H_0$, represents the two most fundamental parameters in observational cosmology~\cite{Sandage1970PhT}.

The deceleration parameter is predicted to decrease with the cosmic expansion according to the cosmological constant dark energy plus cold dark matter ($\Lambda$CDM) model, which achieves  great  success in describing  the cosmic evolution.   However, this $\Lambda$CDM model suffers a pronounced tension with the latest BAO results from the Dark Energy Spectroscopic Instrument (DESI)~\cite{Adame2025JCAP1,Adame2025JCAP2,Adame2025JCAP3}. The DESI BAO Data Release 2 (DR2)~\cite{AbdulKarim2025PhRvD,AbdulKarim2025PhRvD2,Brodzeller2025PhRvD,Andrade2025PhRvD}, when combined with CMB and SNe Ia data, exhibits a deviation from the $\Lambda$CDM model, with a reported significance reaching up to 4.2$\sigma$~\cite{AbdulKarim2025PhRvD}, which  represents the most compelling evidence to date in favor of a dynamical dark energy scenario. This prompts  increased investigation into  the redshift evolution of the deceleration parameter $q(z)$ beyond the standard $\Lambda$CDM model~\cite{Koussour2023EPJP,Koussour2023PDU,Beesham2025arXiv,Mahanta2025PDU,Mendoza-Martinez2024EPJC,Khurana2024PDU,Pawde2024EPJC,Naik2023EPJC}. For example, DESI Collaboration~\cite{Lodha2025PhRvD} conducted an extended analysis of dark energy constraints by combining the DESI DR2 BAO~\cite{AbdulKarim2025PhRvD}, Planck CMB~\cite{PlanckCollaboration2020A&A}, and three SNe Ia compilations (Pantheon+~\cite{Scolnic2022ApJ}, Union3~\cite{Rubin2025ApJ}, and DES5YR~\cite{DESCollaboration2024ApJL}). By employing both parametric and nonparametric methods to construct the evolution of dark energy, they derived the evolution of the deceleration parameter. Their analysis revealed a weaker cosmic acceleration at low redshift compared to the $\Lambda$CDM model. Recently, Wang et al.~\cite{Wang2025EPJC} revisited the evolution of cosmic acceleration in a spatially flat $w_{0}w_{a}$CDM cosmological model by using the same BAO, CMB, and SNe Ia samples. By analyzing both the deceleration parameter $q(z)$ and the jerk parameter $j(z)$, and validating results through the $Om(z)$ diagnostic, they found that the DESI BAO, DESI BAO+CMB, and DESI BAO+CMB+Union3/DES5YR combinations provide  evidence for a slowing down of cosmic acceleration at late times. These results suggest that, within the $w_{0}w_{a}$CDM framework, cosmic acceleration has already reached its maximum and is currently in decline.

In parallel with the model-based analyses summarized above, model-independent reconstruction approaches have been employed to infer $q(z)$ directly from observational data without assuming any specific cosmological model. Such reconstruction can be divided into two classes: parametric and nonparametric. Parametric reconstructions adopt general parametrizations of $q(z)$ and determine their coefficients from data~\cite{Wang2025EPJC,Mamon2017EPJC,Camarena2020PhRvR,Myrzakulov2025NuPhB,Naik2023PhLB}. Nonparametric reconstructions, on the other hand, estimate $q(z)$ from observations using statistical techniques such as Gaussian processes (GP) and Bayesian methods~\cite{Wang2018ApJL,Jesus2020JCAP,Shafieloo2012PhRvD,Xu2020Ap&SS,Mukherjee2020arXiv}. For example, Mamon and Das~\cite{Mamon2017EPJC} proposed a logarithmic parametrization to reconstruct the deceleration parameter $q(z)$. They found that   the best-fit values of $q_0$ and the redshift $z_t$ at which the transition from deceleration to acceleration occurs, derived  from CMB, BAO, and SNe Ia data,  are in good agreement with the $\Lambda$CDM model within $2\sigma$ uncertainties. As a representative nonparametric analysis,   
Jesus et al.~\cite{Jesus2020JCAP} employed GP to reconstruct the deceleration parameter $q(z)$ from the observational Hubble parameter data (OHD) and SNe Ia data respectively, yielding transition redshift estimates of $z_t = 0.59_{-0.11}^{+0.12}$ from the OHD data and $0.683_{-0.082}^{+0.11}$ from the SNe Ia data.  
 
Within the parametric class of reconstructions, Lagrange interpolation~\cite{Berrut2004SIAMR} has recently gained attention in cosmology as a simple yet flexible tool for reconstructing a redshift-dependent function from its values at a finite number of interpolation nodes~\cite{Wang2004ApJ,Wang2006PhLB,Cardenas2015PhLB,Grandon2021CQGra,Bernardo2022PDU,Orchard2024PDU}. For example, in Refs.~\cite{Wang2004ApJ,Wang2006PhLB}, Lagrange interpolation was employed to reconstruct the dark energy density using data from SNe Ia, the CMB shift parameter, and large-scale structure growth. It was found that this method provided tighter constraints on the dark energy density compared to parametrizations of the dark energy equation of state (EOS). Li and Wang~\cite{Li2025EPJC} utilized this Lagrange interpolation method to reconstruct the dark energy EOS $w(z)$ and dark energy density function $f(z)$ with the latest DESI DR2 BAO. They found that current observations favor dynamical dark energy with deviations of the EOS from $w=-1$ at over $2\sigma$, and the reconstructed $f(z)$ exhibits a hump around $z\sim0.5$.

Motivated by the demonstrated advantages of Lagrange interpolation in reconstructing the dark energy evolution, we extend this model-independent technique in this paper to study  the evolution of the deceleration parameter. We will implement both third- and fourth-order Lagrange interpolation in terms of the standard redshift $z$, the compressed redshift $y \equiv z/(1+z)$, and the logarithmic redshift $\zeta \equiv \ln(1+z)$, with the latter two being particularly advantageous for mitigating high-$z$ divergence issues. To achieve more robust results, we plan to jointly constrain the deceleration parameter and the Hubble constant with the latest observational datasets, including DESI BAO, CMB distance priors, and three SNe Ia datasets: DES-Dovekie~\cite{Popovic2026MNRAS}, which is a fully recalibrated reanalysis of the original DES5YR dataset~\cite{DESCollaboration2024ApJL}, Pantheon+, and Union3. Given the possible tension highlighted by Pang et al.~\cite{Pang2025SCPMA} between the $H_0$ prior~\cite{Riess2022ApJL} and the DESI DR2 BAO-favored dynamical dark energy, we also perform parallel analyses with and without the $H_0$ prior to quantify its impact on the reconstruction of $q(z)$. For model selection, we adopt Bayesian inference~\cite{Kass1995JASA,Trotta2008ConPh}, and frequentist method~\cite{Cowan2011EPJC,Wilks1938}.

This paper is organized as follows: Section~\ref{S2} introduces Lagrange interpolation and the methods for model comparison. Section~\ref{S3} details the datasets used in this study, such as BAO, SNe Ia, and CMB data. Section~\ref{S4} presents the constraint results of our analysis, including the best-fitting values of the parameters and their implications for cosmic evolution. Finally, Section~\ref{S5} summarizes our findings and discusses their significance in the context of cosmological studies.


\section{METHODOLOGY}\label{S2}

The deceleration parameter, which plays a central role in describing the expansion history of the Universe, is defined as:
\begin{equation}
q(z) \equiv -\frac{\ddot{a}(z)a(z)}{\dot{a}^{2}(z)} = \frac{(1+z)}{E(z)}\frac{dE(z)}{dz} - 1,
\label{qz-qa}
\end{equation}
where $z = \frac{1}{a}-1$ represents the redshift, and $E(z)=H(z)/H_0$ is the dimensionless Hubble parameter. For the spatially flat $\Lambda$CDM model, the dimensionless Hubble parameter takes the form  $E(z)^2=\left[\Omega_{m,0}(1+z)^3+(1-\Omega_{m,0})\right]$.  From it, one can derive the deceleration parameter
\begin{equation}
q_{\Lambda{\rm CDM}}(z) = \frac{3\,\Omega_{m,0}(1+z)^3} {2\left[\Omega_{m,0}(1+z)^3+(1-\Omega_{m,0})\right]} -1 ,
\label{qLCDM}
\end{equation}
where $\Omega_{m,0}$ is the present matter density parameter, and radiation is neglected.

In this paper, we express $q(z)$ as the sum of a fiducial component and a freely varying deviation term:
\begin{equation}
q(z) = q_{\rm fid}(z) + \delta q(z),
\label{qzdqz}
\end{equation}
where $q_{\rm fid}$ denotes the deceleration parameter predicted by the fiducial model, which in this work is assumed to  be  the spatially flat $\Lambda$CDM model, i.e., $q_{\rm fid}=q_{\Lambda{\rm CDM}}$. The term $\delta q$ represents the deviation of the deceleration parameter between the real Universe and the $\Lambda$CDM model. Rather than directly reconstructing the deceleration parameter $q(z)$, our approach centers on reconstructing its deviation from the $\Lambda$CDM model, $\delta q(z)$, by using the Lagrange interpolation method. This strategy enhances numerical stability, reduces the risk of overfitting, and enables direct consistency checks with the standard cosmological model. By combining the reconstructed $\delta q(z)$ with $q_\mathrm{\Lambda CDM}$, we can obtain $q(z)$, which depends on $\Omega_{m,0}$ and the free parameters introduced in the Lagrange interpolation. Integrating $q(z)$ yields the expression of  $E(z)$, which can then be used to find   $H(z)$ through the relation  $H(z)=H_0 E(z)$.

Since DESI BAO data indicate that the deviation of the $q(z)$ evolution from the $\Lambda$CDM model primarily occurs at late times~\cite{Lodha2025PhRvD} and the DESI BAO sample extends up to $z_{\max}\simeq 2.33$, we adopt a conservative cutoff at $z_c=2.5$. We assume that the cosmic expansion follows the predictions of the $\Lambda$CDM model for $z>z_c$. Therefore, we can set $\delta q(z)=0$ for $z > z_c$. We also consider other cutoffs, such as $z_c=3$ or $5$, and find that the main conclusions derived from $z_c=2.5$ do not change significantly.

\subsection{Lagrange interpolation method}\label{SS2a}

The cosmological parameters to be fit in this method include the values of $\delta q(z_i)$ at $n$ discrete redshifts, where $0\leq i<n$ and $0\leq z_i<2.5$. The values of $\delta q(z)$ between these $n$ nodes are determined by Lagrange interpolation:
\begin{equation}
\delta q(u) = \sum_{i=0}^{n} x_i
\prod_{\substack{j=0 \\ j \neq i}}^{n} \frac{u - u_j}{u_i - u_j}
\end{equation}
with $n$ denoting the order of interpolation. $n =2, 3, 4$ correspond to quadratic, cubic, and quartic interpolation, respectively. Since a quadratic interpolation captures only a single parabolic trend and previous studies~\cite{Wang2018ApJL,Lodha2025PhRvD} indicate that the evolving effective dark energy density may exhibit nontrivial structure such as oscillatory behavior, we therefore reconstruct $\delta q$ using cubic and quartic Lagrange interpolation. The interpolation variable $u$ is chosen to be the standard redshift  $z$, the $y$-redshift $y= z/(1+z)$, and the log-redshift $\zeta=\ln(1+z)$. The latter two are particularly suitable for SNe Ia and BAO data, which are typically denser at low $z$ and sparser at high $z$. The coefficients $x_i = \delta q(u_i)$ represent the deviations evaluated at the interpolation nodes $u_i$, for $i = 0, 1, 2, \ldots, n$. The node $u_0$ corresponds to the present redshift $z_0=0$, while $u_n$ corresponds to the cutoff redshift $z_n=z_{\rm c}$, with the boundary condition $x_n=\delta q(z_{\rm c})=0$. Notably, the interior nodes $u_i$ are chosen such that the corresponding values are uniformly spaced, i.e., $u_1 - u_0 = \ldots = u_{i+1} - u_i = \ldots = u_n - u_{n-1}$. The use of a uniform grid in $u$ is mainly motivated by simplicity and numerical stability. This choice is also consistent with Refs.~\cite{Wang2018ApJL,Grandon2021CQGra,Lodha2025PhRvD,Li2025EPJC}, where equal partitioning is employed to reconstruct the effective dark energy density.

\subsection{Assessment of Lagrange interpolation schemes}\label{SSS2c}

The higher-order Lagrange interpolation provides a more flexible and precise representation of the deceleration parameter compared to the lower-order approach, as it can capture more intricate details of cosmic evolution. However, this increased flexibility introduces additional parameters, raising the question of how many terms are necessary to achieve an optimal balance between accuracy and complexity. To quantitatively evaluate and compare these Lagrange interpolation schemes, we use Bayesian evidence and frequentist methods for model selection.

\subsubsection{Bayesian evidence}
Bayesian evidence is a widely used tool in cosmology for evaluating the performance of cosmographic models. It can achieve a balance between model complexity and goodness-of-fit, helping to avoid overfitting while effectively capturing essential cosmic dynamics.
 For a given model $\mathcal{M}$ with parameter space $\theta$ and observational data $d$, the Bayesian evidence $\tilde{E}$ is expressed as:
\begin{equation} \tilde{E} (d|\mathcal{M}) = \int \mathcal{L} (d|\theta, \mathcal{M}) \pi(\theta|\mathcal{M}) d\theta, \end{equation} where $\pi(\theta|\mathcal{M})$ represents the prior distribution of parameters in model $\mathcal{M}$, and $\mathcal{L}(d|\theta, \mathcal{M})$ is the likelihood of the data given the parameters. The likelihood function can be expressed as  $\mathcal{L} \propto e^{-\chi^2/2}$. The expressions for the $\chi^2$  functions corresponding to different datasets  will be provided in the next section.

When model $\mathcal{M}_i$ is compared with the reference model $\mathcal{M}_j$, the Bayes factor $B_{ij}$, representing the ratio of their evidences, is calculated as:
\begin{equation}
\ln B_{ij} = \ln \tilde{E}(d|\mathcal{M}_i) - \ln \tilde{E} (d|\mathcal{M}_j),
\end{equation}
which allows us to evaluate the strength of evidence for one model over the other. The revised Jeffreys scale by Kass and Raftery~\cite{Kass1995JASA} (Table~\ref{Tabrule}) is used to interpret the significance of the Bayes factor. In our analysis, the reference model $\mathcal{M}_j$ is taken to be the $\Lambda$CDM model.

\begin{table}[h]
 \caption{\label{Tabrule}Revised Jeffreys scale quantifying the strength of evidence for model $\mathcal{M}_i$ compared with model $\mathcal{M}_j$. Negative values of $\ln B_{ij}$ indicate a preference for the reference model $\mathcal{M}_j$, which is chosen to be $\Lambda$CDM in this work.}
 \centering
 \begin{tabular}{c|c}
 \hline
$\ln B_{ij}$ & Strength of evidence for model $\mathcal{M}_i$ \\
\hline
$0 < |\ln B_{ij}| < 1$ & Weak \\
$1 < |\ln B_{ij}| < 3$ & Definite/positive \\
$3 < |\ln B_{ij}| < 5$ & Strong \\
$|\ln B_{ij}| > 5$ & Very strong \\
\hline
\end{tabular}
\end{table}

By applying Bayesian evidence, we can compare different Lagrange interpolation schemes to identify the model that best fits the observational data. 

\subsubsection{Frequentist methods}

Frequentist methods~\cite{Cowan2011EPJC} are a class of statistical techniques based on the principles of frequentist statistics, which is one of the main approaches to statistical inference. We use them to measure the statistical significance of the preference for the reconstructed model over the $\Lambda$CDM model from a given data combination. Our measures are based on
\begin{equation}
\label{eq:dchi2min}
\Delta\chi^2_{\rm min} \equiv -2\,\Delta\ln\mathcal{L}_{\rm max},
\end{equation}
where $\chi^2_{\rm min}$ is the minimum value of $\chi^2$,  $\mathcal{L}_{\rm max}$ represents the maximum likelihood, and $\Delta$ means the difference between the reconstructed model and the $\Lambda$CDM model. Following Wilks' theorem~\cite{Wilks1938}, we can translate $\Delta\chi^2_{\rm min}$ into frequentist significance $N\sigma$ by using
\begin{equation}
\label{eq:frequentist}
{\rm CDF}_{\chi^2}\!\left(-\Delta\chi^2_{\rm min}\,|\,\Delta \nu~{\rm dof}\right)=\frac{1}{\sqrt{2\pi}} \int_{-N}^{N} e^{-t^2/2}\,{\rm d}t ,
\end{equation}
where CDF denotes the cumulative distribution function of the $\chi^2$ distribution, and $\Delta\nu=3$ and $4$ for the cubic and quartic reconstructions, respectively.

\section{Datasets}\label{S3}

Our study utilizes the latest observational datasets to explore the evolution of the Universe. These datasets encompass SNe Ia, BAO, and distance priors derived from CMB measurements, offering comprehensive constraints on cosmological parameters.

\subsection{SNe Ia}
Since it was shown in Ref.~\cite{AbdulKarim2025PhRvD} that different SNe Ia datasets, when combined with CMB and DESI BAO, exhibit slightly different preferences for dynamical dark energy, we include three different SNe Ia compilations in our analysis. First, we use the Pantheon+ sample~\cite{Scolnic2022ApJ}, a compilation of 1701 light curves involving 1550 spectroscopically confirmed SNe Ia spanning a redshift range from 0.001 to 2.26, which provides a comprehensive view of the cosmic expansion history. Notably, due to the high sensitivity of SNe Ia to peculiar velocities at $z < 0.01$, the Hubble residual bias is non-negligible. Consequently, we restrict our analysis to SNe Ia data with $z > 0.01$ instead of the full sample to mitigate this effect. For a more detailed discussion, we refer the reader to Ref.~\cite{Brout2022ApJ}. In addition, we include the Union3 compilation~\cite{Rubin2025ApJ}, which consists of 2087 SNe Ia over the redshift interval $0.01 < z < 2.26$, processed through the Unity 1.5 pipeline based on Bayesian Hierarchical Modeling. Finally, we incorporate the DES-Dovekie dataset~\cite{Popovic2026MNRAS}, a fully recalibrated sample obtained from a reanalysis of the original DES5YR dataset~\cite{DESCollaboration2024ApJL}, which includes an improved photometric cross-calibration, recent white-dwarf observations to cross-calibrate DES with low-redshift surveys, a retraining of the SALT3 light-curve model, and a correction to a numerical approximation in the host-galaxy color law. It comprises 201 low-redshift SNe Ia, along with 1619 photometrically classified SNe Ia. This combination allows for a comprehensive and cross-validated reconstruction of the cosmic expansion history using independent light-curve training methodologies and redshift distributions.

The observed distance modulus, $\mu_{\mathrm{obs} }$, is usually defined as
\begin{equation}
\mu_{\mathrm{obs} }=m_{\mathrm{B} }^{*}+\alpha X_{1}-\beta C-M_{\mathrm{B} } \;,
\end{equation}
where $m_{\mathrm{B} }^{* }$ denotes the observed peak magnitude in the rest-frame B band, $X_1$ characterizes the stretch of the light curve, $C$ is the color of the SNe Ia at maximum brightness, $M_{\mathrm{B}}$ is the absolute magnitude, and $\alpha$ and $\beta$ are nuisance parameters in the distance estimate. For Pantheon+ and DES-Dovekie,  which utilize the BEAMS with Bias Corrections method~\cite{Kessler2017ApJ} to calibrate the SNe Ia, the observed corrected apparent magnitude is given by $m_{B,\mathrm{corr}}^{*}=m_{B}^{*}+\alpha X_{1}-\beta C+\Delta_{B}$, where $\Delta_{B}$ denotes the bias-correction term~\cite{Popovic2021ApJ}. For Union3, the binned supernova distances are obtained from the UNITY1.5 framework, and we refer the reader to Ref.~\cite{Rubin2025ApJ} for details. Then, the observed distance modulus can be reexpressed as
\begin{equation}
\mu_{\mathrm{obs}}=m_{B,\mathrm{corr}}^{*}-M_{B}.
\end{equation}
 The theoretical distance modulus is given by
\begin{equation}
\mu_{\mathrm{th}}\left ( z \right ) =5\log\left ( d_{\mathrm{L} } \right ) +\mu_{0} \;,
\end{equation}
where 
\begin{equation}
d_{\mathrm{L}}(z)=(1+z)\int_{0}^{z}\frac{dz'}{E(z')}
\end{equation}
is the Hubble-constant free luminosity distance, and $\mu_{0}=42.38-5\log{h}$ with $h\equiv H_{0}/100\,\mathrm{km\,s^{-1}\,Mpc^{-1}}$. The constraints on cosmological parameters from the SNe Ia data can be obtained by minimizing the following $\chi^{2}$ 
\begin{equation}
\chi _{\mathrm{SNeIa} }^{2} = \Delta \mu^{T}\cdot (C_{\mathrm{SNeIa}})^{-1}\cdot\Delta \mu ,
\end{equation}
where $\Delta \mu \equiv m^{*}_{B,\mathrm{corr}}-5\log d_{\mathrm{L} }-(M_{B}+\mu_{0})$, and $C_{\mathrm{SNeIa}}$ is the covariance matrix of the SNe Ia data. Following the method proposed by Conley et al. in Ref.~\cite{Conley2011ApJS}, the quantity $M_{B}+\mu_{0}$ is analytically marginalized over.
Consequently, the $\chi^2_\mathrm{SNeIa}$ function can be rewritten as
\begin{equation}
\chi^{2}_{\mathrm{SNeIa}}=a+\ln{\frac{f}{2\pi } }-\frac{b^2}{f}, 
\end{equation}
where $a\equiv (\Delta m )^{T}\cdot (C_{\mathrm{SNeIa}})^{-1}\cdot \Delta m$, $b\equiv (\Delta m)^{T}\cdot (C_{\mathrm{SNeIa}})^{-1}\cdot \mathbf{1}$, $f\equiv \mathbf{1}^T\cdot (C_{\mathrm{SNeIa}})^{-1}\cdot  \mathbf{1}$, and $\Delta m = m^{*}_{B,\mathrm{corr}}-5\log d_{\rm L}$.

\subsection{BAO}
The BAO dataset used in this study incorporates the latest measurements from DESI DR2~\cite{AbdulKarim2025PhRvD}, which marks a substantial expansion over DESI DR1. The DESI survey spans three optical bands (\textit{g}, \textit{r}, and \textit{z}), which are used to select four classes of extragalactic targets: bright galaxies, luminous red galaxies, emission line galaxies, and quasars. Collected with the DESI instrument on the Mayall Telescope, DR2 includes BAO measurements from over 13.1 million galaxies and 1.6 million quasars, covering the redshift range $0.295 \leq z \leq 2.330$. The BAO observables~\cite{Eisenstein2005ApJ} used in this work involve three types of distance ratios: $D_{\mathrm{M}}(z)/r_{\mathrm{d}}$, $D_{\mathrm{H}}(z)/r_{\mathrm{d}}$, and $D_{\mathrm{V}}(z)/r_{\mathrm{d}}$. Here, $D_{\mathrm{M} }$ denotes the transverse comoving distance, $D_{\mathrm{H} }$ is the Hubble distance, and $D_{\mathrm{V} }$ is the volume-averaged distance,  which are defined as follows: 
\begin{equation}
\begin{aligned}
D_{\mathrm{M}}(z) &= (1+z)D_{\mathrm{A}}(z), \\
D_{\mathrm{H}}(z) &= \frac{c}{H(z)},   \\
D_{\mathrm{V}}(z) &= \left[z\,D_{\mathrm{M}}^{2}(z)\,D_{\mathrm{H}}(z)\right]^{1/3},
\end{aligned}
\end{equation}
with $D_{\mathrm{A}}(z)=\frac{c}{1+z}\int_{0}^{z}\frac{dz'}{H(z')}$ being the angular diameter distance in a flat universe.
The quantity $r_{\mathrm{d}} \equiv r_{s}(z_{d})$ is the comoving sound horizon at the baryon drag epoch $z_{d}$, which sets the standard ruler for BAO measurements. The comoving sound horizon at redshift $z$ is defined by $r_{s}(z)=\int_{z}^{\infty}\frac{c_{s}(z')}{H(z')}\,dz'$, where $c_{s}(z)$ is the sound speed in the photon-baryon fluid, given by
\begin{equation}
\label{soundspeed}
c_{s}(z)=\frac{c}{\sqrt{3\left ( 1+\frac{3\rho_{b,0} }{4\rho_{\gamma,0}(1+z) } \right ) } }.
\end{equation}
Here, $\rho_{b,0}$ and $\rho_{\gamma,0}$ denote the present-day energy densities of baryons and photons, respectively, and satisfy the relation $3\rho_{b,0} /(4\rho_{\gamma,0}) =31500\Omega_{b,0}h^{2}\left ( T_{\mathrm{CMB} }/2.7\mathrm{K} \right )^{-4}.$ The value of $z_{\mathrm{d}}$ is determined using the fitting formula given in Ref.~\cite{Eisenstein1998ApJ}, as follows:
\begin{equation}
z_{\mathrm{d}}
=
\frac{1291\left(\Omega_{m,0}h^{2}\right)^{0.251}}
{1+0.659\left(\Omega_{m,0}h^{2}\right)^{0.828}}
\left[
1+b_{1}\left(\Omega_{b,0}h^{2}\right)^{b_{2}}
\right],
\label{eq:zd}
\end{equation}
where
\begin{equation*}
\begin{aligned}
b_{1} &= 0.313\left(\Omega_{m,0}h^{2}\right)^{-0.419}
\left[1+0.607\left(\Omega_{m,0}h^{2}\right)^{0.674}\right], \\
b_{2} &= 0.238\left(\Omega_{m,0}h^{2}\right)^{0.223}.
\end{aligned}
\end{equation*}
Here,  $\Omega_{b,0}$ denotes the present-day baryon density parameter.
The $\chi^{2}$ function for each BAO dataset can be written as:
\begin{equation}
\chi^2_{\mathrm{BAO}}
=
(\bold{A}^{\mathrm{th}}-\bold{A}^{\mathrm{obs}})^{T}
\,\mathrm{\bold{Cov}}^{-1}\,
(\bold{A}^{\mathrm{th}}-\bold{A}^{\mathrm{obs}}).
\end{equation}
Here, $\bold{A}^{\mathrm{obs}}$ is the vector consisting of the observed values of $D_M/r_d$, $D_H/r_d$, and $D_V/r_d$, $\bold{A}^{\mathrm{th}}$ is the vector of theoretical predictions, and $\mathrm{\bold{Cov}}$ is the  $13\times13$ covariance matrix of the BAO dataset.

\subsection{CMB distance prior}
The CMB distance prior is characterized by three cosmological parameters: the acoustic scale
\begin{equation}
l_A \equiv (1+z_*)\frac{\pi D_{\mathrm{A}}(z_*)}{r_s(z_*)},
\label{eq:lA}
\end{equation}
the shift parameter
\begin{equation}
R \equiv \frac{(1+z_*)\sqrt{\Omega_{m,0}}\,H_0\, D_{\mathrm{A}}(z_*)}{c},
\label{eq:R}
\end{equation}
and $\Omega_{b,0} h^2$.
Here $z_*$ is the redshift of photon decoupling, which is calculated using the fitting formula given in Ref.~\cite{Hu1996ApJ}:
\begin{equation*}
z_*=1048\left[1+0.00124\left(\Omega_{b,0}h^2\right)^{-0.738}\right]\left[1+g_1\left(\Omega_{m,0}h^2\right)^{g_2}\right],
\end{equation*}
where
\begin{equation*}
\begin{aligned}
g_1 &=\frac{0.0783\left(\Omega_{b,0}h^2\right)^{-0.238}}{1+39.5\left(\Omega_{b,0}h^2\right)^{0.763}},\\
g_2 &=\frac{0.560}{1+21.1\left(\Omega_{b,0}h^2\right)^{1.81}}.
\end{aligned}
\end{equation*}
According to the Planck 2018 measurements, these parameter values are $l_A = 301.471^{+0.089}_{-0.090}$, $R = 1.7502 \pm 0.0046$, and $\Omega_{b,0} h^2 = 0.02236 \pm 0.00015$~\cite{Chen2019JCAP}, which can serve as a reliable substitute for the full dataset published by the Planck satellite in 2018. 
The $\chi^{2}$  for the CMB distance priors can be expressed as
\begin{equation}
\chi _{\mathrm{CMB} }^{2} = \sum\limits_{i,j} \left ( x_{i}^{\mathrm{obs} }-x_{i}^{\mathrm{th} } \right ) \left ( C_{\mathrm{CMB} }^{-1} \right )_{ij} \left ( x_{j}^{\mathrm{obs} }-x_{j}^{\mathrm{th} } \right ),
\end{equation}
where $x=\left \{ R, l_{\mathrm{A} }, \Omega_{b,0} h^2 \right \}$. The superscripts $\mathrm{obs}$ and $\mathrm{th}$  denote the values derived from the CMB observations and the corresponding theoretical values, respectively. $C_{\mathrm{CMB}}^{-1}$ is  the inverse of the covariance matrix, given as follows~\cite{Chen2019JCAP}
\begin{equation}
C_{\mathrm{CMB}}^{-1}=
\begin{pmatrix}
 94392.3971 &  -1360.4913 & 1664517.2916 \\
  -1360.4913 &    161.4349 &    3671.6180 \\
 1664517.2916 &    3671.6180 & 79719182.5162
\end{pmatrix}.
\end{equation}

\subsection{Local measurement of $H_0$}
We adopt the local determination of the Hubble constant from the Supernovae and $H_0$ for the Equation of State (SH0ES) Collaboration, $H_0 = 73.04 \pm 1.04 \, \mathrm{km\,s^{-1}\,Mpc^{-1}}$, derived from the calibration of the Cepheid variable distance scale. This calibration was based on geometrically anchored Cepheid measurements from Gaia EDR3 parallaxes, masers in NGC 4258, and detached eclipsing binaries in the Large Magellanic Cloud, as reported by Riess et al.~\cite{Riess2022ApJL}. The $\chi^{2}$  for the Hubble constant is  \begin{equation}
\chi^{2}_{H_0}=\left(\frac{H_0^{\mathrm{th}}-73.04}{1.04} \right)^{2}.
\end{equation}

The constraints on free parameters from all observational data considered in the present paper can be obtained by minimizing the following $\chi_{\mathrm{total}}^2$
\begin{eqnarray}
\chi_{\mathrm {total}}^2=\chi_{\mathrm{SNeIa}}^2+\chi_{\mathrm{BAO}}^2+\chi_{\mathrm{CMB}}^2+\chi_{{H_0}}^2.
\end{eqnarray}
We use the publicly available Markov Chain Monte Carlo (MCMC) package \texttt{CosmoMC}~\cite{Lewis2002PhRvD}, combined with the nested sampling plugin PolyChord~\cite{Handley2015MNRAS1,Handley2015MNRAS2} for parameter estimation. The Lagrange interpolation models involve $3+n$ free parameters: $H_0$, $\Omega_{\mathrm{m,0}} h^2$, $\Omega_{\mathrm{b,0}} h^2$, and the set $x_i$ for $i = 0, 1, 2, \ldots, n-1$. We adopt uniform priors over the following ranges: $H_0 \in [40,\, 100]\,\mathrm{km\,s^{-1}\,Mpc^{-1}}$, $\Omega_{\mathrm{m,0}} h^2 \in [0.01,\, 0.5]$, $\Omega_{\mathrm{b,0}} h^2 \in [0.005,\, 0.1]$, and $x_i \in [-1,\, 1]$ for each $i = 0, 1, 2, \ldots, n-1$.

\section{Analyses and results}\label{S4}

\begin{table*}[t]
\caption{Summary of constraints on the cosmological parameters at the $68\%$ CL.}
\label{tab:parameters}
\centering
\scriptsize
\setlength{\tabcolsep}{3pt}
\renewcommand{\arraystretch}{1.15}

\begin{ruledtabular}
\begin{tabular}{llcccccc}
\textbf{Models} & \textbf{Data set} & $H_0$ & $\Omega_{m,0}h^2$ & $\Omega_{b,0}h^2$ & $q_0$ & $q_{\rm pivot}$ & $z_t$ \\
\hline

\multirow{5}{*}{$\Lambda$CDM}
& BC & $68.8\pm0.3$ & $0.1417\pm0.0006$ & $0.0226\pm0.0001$ & $-0.55\pm0.01$ & $0.016\pm0.006$ & $0.67\pm0.01$ \\
& Pantheon+ & $\cdots$ & $\cdots$ & $\cdots$ & $-0.48\pm0.03$ & $\cdots$ & $\cdots$ \\
& BC+Pantheon+ & $68.7\pm0.3$ & $0.1420\pm0.0006$ & $0.0225\pm0.0001$ & $-0.55\pm0.01$ & $0.019\pm0.006$ & $0.67\pm0.01$ \\
& BC+DES-Dovekie & $68.7\pm0.3$ & $0.1420\pm0.0006$ & $0.0225\pm0.0001$ & $-0.55\pm0.01$ & $0.018\pm0.006$ & $0.67\pm0.01$ \\
& BC+Union3 & $68.8\pm0.3$ & $0.1419\pm0.0006$ & $0.0225\pm0.0001$ & $-0.55\pm0.01$ & $0.017\pm0.006$ & $0.67\pm0.01$ \\

\noalign{\vskip 2pt}\hline\noalign{\vskip 1pt}

\multirow{5}{*}{Cubic-$z$}
& BC & $64.2^{+1.7}_{-2.0}$ & $0.1433\pm0.0010$ & $0.0224\pm0.0001$ & $-0.01\pm0.20$ & $-0.083\pm0.040$ & $0.92^{+0.12}_{-0.09}$ \\
& Pantheon+ & $\cdots$ & $\cdots$ & $\cdots$ & $-0.36\pm0.10$ & $\cdots$ & $\cdots$ \\
& BC+Pantheon+ & $67.3\pm0.6$ & $0.1428\pm0.0009$ & $0.0225\pm0.0001$ & $-0.35\pm0.06$ & $-0.034\pm0.026$ & $0.77^{+0.06}_{-0.05}$ \\
& BC+DES-Dovekie & $67.7\pm0.5$ & $0.1427\pm0.0009$ & $0.0225\pm0.0001$ & $-0.38\pm0.06$ & $-0.030\pm0.026$ & $0.76\pm0.05$ \\
& BC+Union3 & $66.3\pm0.8$ & $0.1430\pm0.0009$ & $0.0225\pm0.0001$ & $-0.24\pm0.09$ & $-0.049\pm0.029$ & $0.82\pm0.07$ \\

\noalign{\vskip 2pt}\hline\noalign{\vskip 1pt}

\multirow{5}{*}{Cubic-$y$}
& BC & $64.5^{+1.3}_{-2.6}$ & $0.1428\pm0.0009$ & $0.0225\pm0.0001$ & $0.12^{+0.43}_{-0.17}$ & $-0.037^{+0.019}_{-0.032}$ & $0.77^{+0.06}_{-0.03}$ \\
& Pantheon+ & $\cdots$ & $\cdots$ & $\cdots$ & $-0.24\pm0.22$ & $\cdots$ & $\cdots$ \\
& BC+Pantheon+ & $67.2\pm0.6$ & $0.1426\pm0.0009$ & $0.0225\pm0.0001$ & $-0.31\pm0.12$ & $-0.012\pm0.017$ & $0.72^{+0.04}_{-0.03}$ \\
& BC+DES-Dovekie & $67.5\pm0.6$ & $0.1425\pm0.0009$ & $0.0225\pm0.0001$ & $-0.31\pm0.13$ & $-0.013\pm0.017$ & $0.72^{+0.03}_{-0.03}$ \\
& BC+Union3 & $66.0\pm0.9$ & $0.1426\pm0.0009$ & $0.0225\pm0.0001$ & $-0.09\pm0.20$ & $-0.026\pm0.021$ & $0.75^{+0.04}_{-0.03}$ \\

\noalign{\vskip 2pt}\hline\noalign{\vskip 1pt}

\multirow{5}{*}{Cubic-$\zeta$}
& BC & $63.3^{+1.2}_{-2.2}$ & $0.1431\pm0.0009$ & $0.0225\pm0.0001$ & $0.23^{+0.32}_{-0.13}$ & $-0.083^{+0.026}_{-0.042}$ & $0.86^{+0.08}_{-0.04}$ \\
& Pantheon+ & $\cdots$ & $\cdots$ & $\cdots$ & $-0.32\pm0.15$ & $\cdots$ & $\cdots$ \\
& BC+Pantheon+ & $67.3\pm0.6$ & $0.1426\pm0.0009$ & $0.0225\pm0.0001$ & $-0.31\pm0.08$ & $-0.022\pm0.023$ & $0.74^{+0.05}_{-0.04}$ \\
& BC+DES-Dovekie & $67.6\pm0.6$ & $0.1426\pm0.0009$ & $0.0225\pm0.0001$ & $-0.35\pm0.09$ & $-0.022\pm0.023$ & $0.74^{+0.05}_{-0.04}$ \\
& BC+Union3 & $66.1\pm0.9$ & $0.1428\pm0.0009$ & $0.0225\pm0.0001$ & $-0.15\pm0.13$ & $-0.040\pm0.027$ & $0.77^{+0.05}_{-0.04}$ \\

\noalign{\vskip 2pt}\hline\noalign{\vskip 1pt}

\multirow{5}{*}{Quartic-$z$}
& BC & $65.5^{+2.5}_{-4.2}$ & $0.1433\pm0.0010$ & $0.0224\pm0.0001$ & $-0.19^{+0.62}_{-0.36}$ & $-0.061^{+0.055}_{-0.075}$ & $0.90^{+0.19}_{-0.13}$ \\
& Pantheon+ & $\cdots$ & $\cdots$ & $\cdots$ & $-0.34\pm0.15$ & $\cdots$ & $\cdots$ \\
& BC+Pantheon+ & $67.4\pm0.6$ & $0.1432\pm0.0010$ & $0.0224\pm0.0001$ & $-0.43\pm0.09$ & $-0.033\pm0.026$ & $0.84^{+0.09}_{-0.15}$ \\
& BC+DES-Dovekie & $67.9\pm0.6$ & $0.1430\pm0.0010$ & $0.0225\pm0.0001$ & $-0.45\pm0.10$ & $-0.029\pm0.026$ & $0.79^{+0.07}_{-0.10}$ \\
& BC+Union3 & $66.6\pm1.0$ & $0.1432\pm0.0010$ & $0.0225\pm0.0001$ & $-0.32\pm0.15$ & $-0.045\pm0.030$ & $0.86^{+0.09}_{-0.13}$ \\

\noalign{\vskip 2pt}\hline\noalign{\vskip 1pt}

\multirow{5}{*}{Quartic-$y$}
& BC & $64.4^{+2.8}_{-3.1}$ & $0.1431\pm0.0009$ & $0.0225\pm0.0001$ & $-0.41^{+0.87}_{-0.47}$ & $-0.126\pm0.058$ & $0.92^{+0.09}_{-0.06}$ \\
& Pantheon+ & $\cdots$ & $\cdots$ & $\cdots$ & $0.15\pm0.36$ & $\cdots$ & $\cdots$ \\
& BC+Pantheon+ & $67.3\pm0.6$ & $0.1426\pm0.0009$ & $0.0225\pm0.0001$ & $-0.39\pm0.25$ & $-0.026\pm0.039$ & $0.75\pm0.07$ \\
& BC+DES-Dovekie & $67.5\pm0.7$ & $0.1426\pm0.0009$ & $0.0225\pm0.0001$ & $-0.30\pm0.30$ & $-0.011\pm0.041$ & $0.72\pm0.07$ \\
& BC+Union3 & $66.5\pm1.1$ & $0.1427\pm0.0009$ & $0.0225\pm0.0001$ & $-0.38\pm0.40$ & $-0.053\pm0.041$ & $0.80^{+0.08}_{-0.07}$ \\

\noalign{\vskip 2pt}\hline\noalign{\vskip 1pt}

\multirow{5}{*}{Quartic-$\zeta$}
& BC & $65.9^{+2.8}_{-4.0}$ & $0.1432\pm0.0009$ & $0.0224\pm0.0001$ & $-0.37^{+0.83}_{-0.42}$ & $-0.079\pm0.047$ & $0.93^{+0.14}_{-0.10}$ \\
& Pantheon+ & $\cdots$ & $\cdots$ & $\cdots$ & $-0.23\pm0.24$ & $\cdots$ & $\cdots$ \\
& BC+Pantheon+ & $67.4\pm0.6$ & $0.1429\pm0.0009$ & $0.0225\pm0.0001$ & $-0.47\pm0.15$ & $-0.049\pm0.033$ & $0.83^{+0.08}_{-0.10}$ \\
& BC+DES-Dovekie & $67.8\pm0.6$ & $0.1427\pm0.0009$ & $0.0225\pm0.0001$ & $-0.44\pm0.16$ & $-0.034\pm0.033$ & $0.78^{+0.06}_{-0.09}$ \\
& BC+Union3 & $66.6\pm1.0$ & $0.1430\pm0.0009$ & $0.0225\pm0.0001$ & $-0.35\pm0.25$ & $-0.058\pm0.032$ & $0.85^{+0.09}_{-0.10}$ \\

\end{tabular}
\end{ruledtabular}

\end{table*}

\begin{table*}[t]
\caption{Summary of the $68\%$ CL constraints on the cosmological parameters obtained from the BC+SNe Ia data combinations after imposing the SH0ES $H_0$ prior.}
\label{tab:parameters_H0}
\centering
\scriptsize
\setlength{\tabcolsep}{3pt}
\renewcommand{\arraystretch}{1.15}

\begin{ruledtabular}
\begin{tabular}{llcccccc}
\textbf{Models} & \textbf{Data set} & $H_0$ & $\Omega_{m,0}h^2$ & $\Omega_{b,0}h^2$ & $q_0$ & $q_{\rm pivot}$ & $z_t$ \\
\hline

\multirow{3}{*}{$\Lambda$CDM}
& BC+DES-Dovekie+$H_0$ & $69.0\pm0.3$ & $0.1415\pm0.0006$ & $0.0226\pm0.0001$ & $-0.55\pm0.01$ & $0.013\pm0.005$ & $0.68\pm0.01$ \\
& BC+Pantheon++$H_0$ & $69.0\pm0.3$ & $0.1415\pm0.0006$ & $0.0226\pm0.0001$ & $-0.55\pm0.01$ & $0.013\pm0.006$ & $0.68\pm0.01$ \\
& BC+Union3+$H_0$ & $69.1\pm0.3$ & $0.1414\pm0.0006$ & $0.0226\pm0.0001$ & $-0.56\pm0.01$ & $0.011\pm0.006$ & $0.68\pm0.01$ \\

\noalign{\vskip 2pt}\hline\noalign{\vskip 1pt}

\multirow{3}{*}{Cubic-$z$}
& BC+DES-Dovekie+$H_0$ & $68.8\pm0.5$ & $0.1427\pm0.0009$ & $0.0225\pm0.0001$ & $-0.46\pm0.06$ & $-0.036\pm0.027$ & $0.77\pm0.05$ \\
& BC+Pantheon++$H_0$ & $68.7\pm0.5$ & $0.1428\pm0.0009$ & $0.0225\pm0.0001$ & $-0.43\pm0.05$ & $-0.045\pm0.026$ & $0.79\pm0.05$ \\
& BC+Union3+$H_0$ & $69.0\pm0.7$ & $0.1427\pm0.0009$ & $0.0225\pm0.0001$ & $-0.48\pm0.08$ & $-0.032\pm0.028$ & $0.76\pm0.05$ \\

\noalign{\vskip 2pt}\hline\noalign{\vskip 1pt}

\multirow{3}{*}{Cubic-$y$}
& BC+DES-Dovekie+$H_0$ & $68.8\pm0.5$ & $0.1426\pm0.0009$ & $0.0225\pm0.0001$ & $-0.48\pm0.13$ & $-0.014\pm0.017$ & $0.72^{+0.03}_{-0.03}$ \\
& BC+Pantheon++$H_0$ & $68.7\pm0.5$ & $0.1426\pm0.0009$ & $0.0225\pm0.0001$ & $-0.43\pm0.11$ & $-0.018\pm0.017$ & $0.73^{+0.03}_{-0.03}$ \\
& BC+Union3+$H_0$ & $69.4\pm0.7$ & $0.1425\pm0.0009$ & $0.0225\pm0.0001$ & $-0.62\pm0.17$ & $-0.001\pm0.020$ & $0.70^{+0.04}_{-0.03}$ \\

\noalign{\vskip 2pt}\hline\noalign{\vskip 1pt}

\multirow{3}{*}{Cubic-$\zeta$}
& BC+DES-Dovekie+$H_0$ & $68.8\pm0.5$ & $0.1426\pm0.0009$ & $0.0225\pm0.0001$ & $-0.45\pm0.08$ & $-0.023\pm0.024$ & $0.74^{+0.04}_{-0.04}$ \\
& BC+Pantheon++$H_0$ & $68.7\pm0.5$ & $0.1426\pm0.0009$ & $0.0225\pm0.0001$ & $-0.42\pm0.08$ & $-0.030\pm0.023$ & $0.75^{+0.04}_{-0.03}$ \\
& BC+Union3+$H_0$ & $69.1\pm0.7$ & $0.1426\pm0.0009$ & $0.0225\pm0.0001$ & $-0.50\pm0.11$ & $-0.014\pm0.026$ & $0.72^{+0.05}_{-0.04}$ \\

\noalign{\vskip 2pt}\hline\noalign{\vskip 1pt}

\multirow{3}{*}{Quartic-$z$}
& BC+DES-Dovekie+$H_0$ & $69.1\pm0.5$ & $0.1434\pm0.0010$ & $0.0225\pm0.0001$ & $-0.59\pm0.10$ & $-0.033\pm0.026$ & $0.85^{+0.09}_{-0.15}$ \\
& BC+Pantheon++$H_0$ & $68.8\pm0.5$ & $0.1434\pm0.0010$ & $0.0225\pm0.0001$ & $-0.54\pm0.09$ & $-0.042\pm0.026$ & $0.88^{+0.09}_{-0.16}$ \\
& BC+Union3+$H_0$ & $69.6\pm0.7$ & $0.1435\pm0.0010$ & $0.0225\pm0.0001$ & $-0.70\pm0.12$ & $-0.019\pm0.029$ & $0.87^{+0.15}_{-0.25}$ \\

\noalign{\vskip 2pt}\hline\noalign{\vskip 1pt}

\multirow{3}{*}{Quartic-$y$}
& BC+DES-Dovekie+$H_0$ & $69.2\pm0.6$ & $0.1427\pm0.0009$ & $0.0225\pm0.0001$ & $-0.81\pm0.27$ & $-0.062\pm0.039$ & $0.81^{+0.07}_{-0.06}$ \\
& BC+Pantheon++$H_0$ & $68.9\pm0.6$ & $0.1427\pm0.0009$ & $0.0225\pm0.0001$ & $-0.69\pm0.24$ & $-0.062\pm0.038$ & $0.81^{+0.07}_{-0.06}$ \\
& BC+Union3+$H_0$ & $70.0\pm0.7$ & $0.1428\pm0.0009$ & $0.0225\pm0.0001$ & $-1.26^{+0.11}_{-0.28}$ & $-0.088\pm0.037$ & $0.87^{+0.06}_{-0.04}$ \\

\noalign{\vskip 2pt}\hline\noalign{\vskip 1pt}

\multirow{3}{*}{Quartic-$\zeta$}
& BC+DES-Dovekie+$H_0$ & $69.2\pm0.5$ & $0.1430\pm0.0009$ & $0.0225\pm0.0001$ & $-0.68\pm0.16$ & $-0.057\pm0.031$ & $0.85^{+0.09}_{-0.10}$ \\
& BC+Pantheon++$H_0$ & $68.9\pm0.5$ & $0.1431\pm0.0009$ & $0.0225\pm0.0001$ & $-0.63\pm0.14$ & $-0.069\pm0.032$ & $0.88\pm0.09$ \\
& BC+Union3+$H_0$ & $69.9\pm0.8$ & $0.1432\pm0.0009$ & $0.0225\pm0.0001$ & $-0.96\pm0.20$ & $-0.065\pm0.032$ & $0.93^{+0.12}_{-0.09}$ \\

\end{tabular}
\end{ruledtabular}
\end{table*}

\begin{figure*}[t]
\centering
\begin{minipage}{0.32\textwidth}
  \centering
  \includegraphics[width=\textwidth]{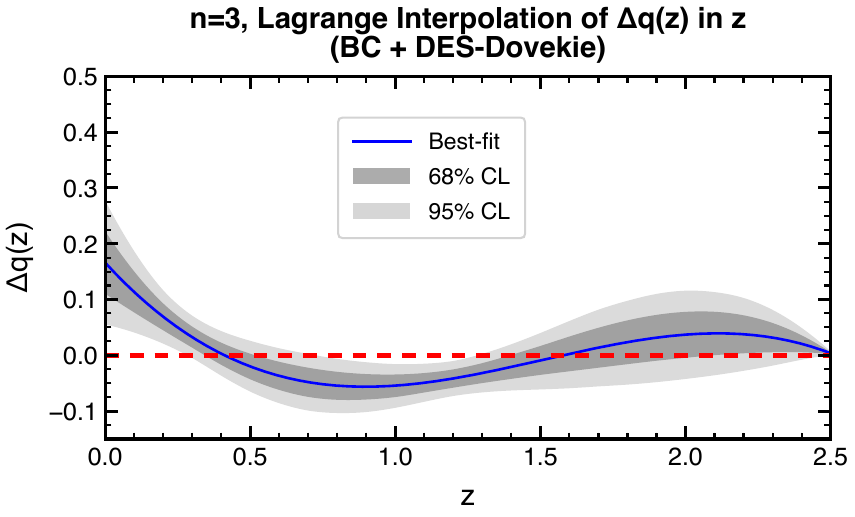}
\end{minipage}\hfill
\begin{minipage}{0.32\textwidth}
  \centering
  \includegraphics[width=\textwidth]{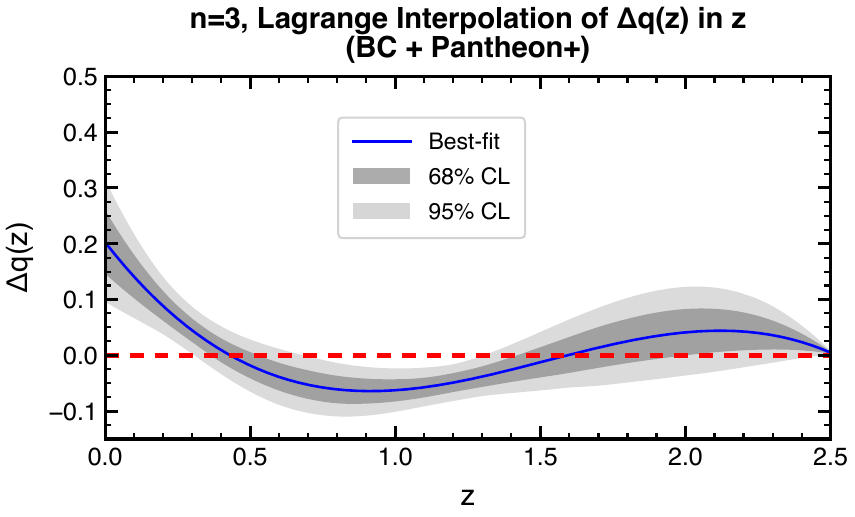}
\end{minipage}\hfill
\begin{minipage}{0.32\textwidth}
  \centering
  \includegraphics[width=\textwidth]{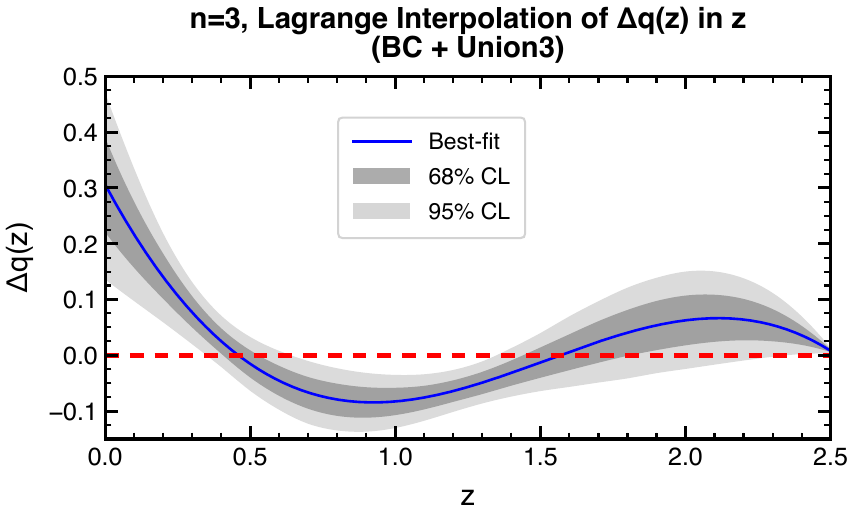}
\end{minipage}

\vspace{1ex} 
\begin{minipage}{0.32\textwidth}
  \centering
  \includegraphics[width=\textwidth]{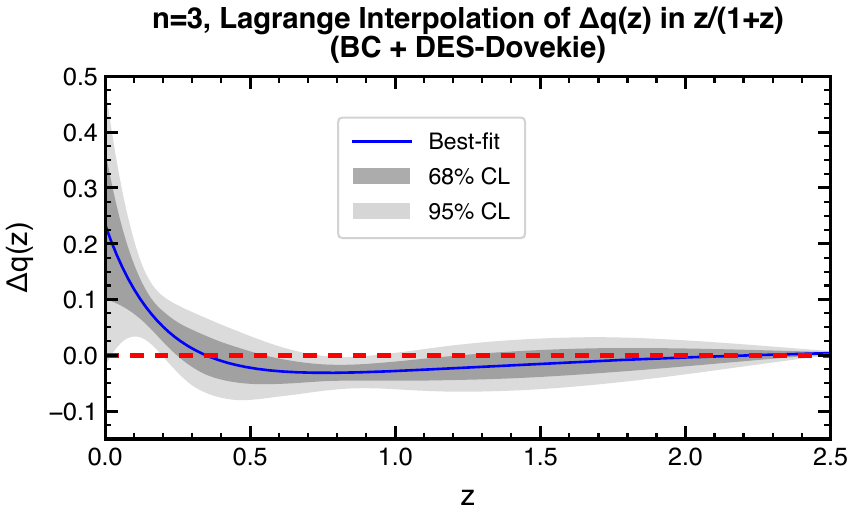}
\end{minipage}\hfill
\begin{minipage}{0.32\textwidth}
  \centering
  \includegraphics[width=\textwidth]{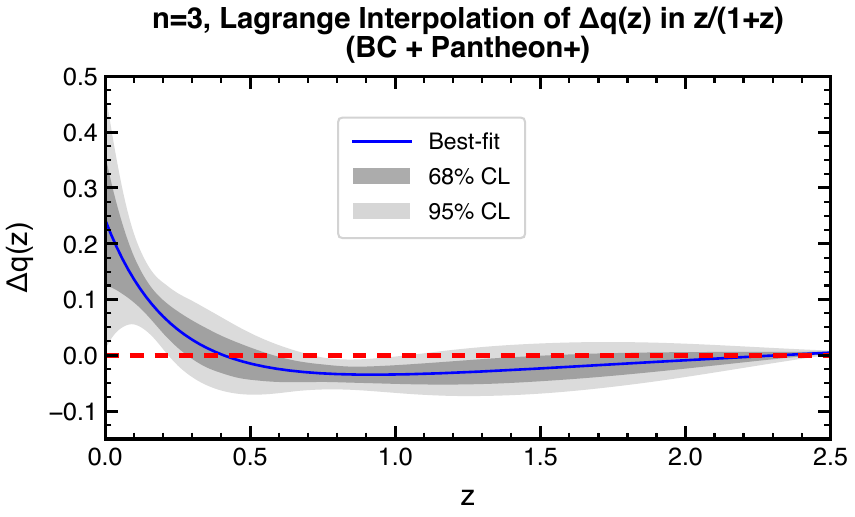}
\end{minipage}\hfill
\begin{minipage}{0.32\textwidth}
  \centering
  \includegraphics[width=\textwidth]{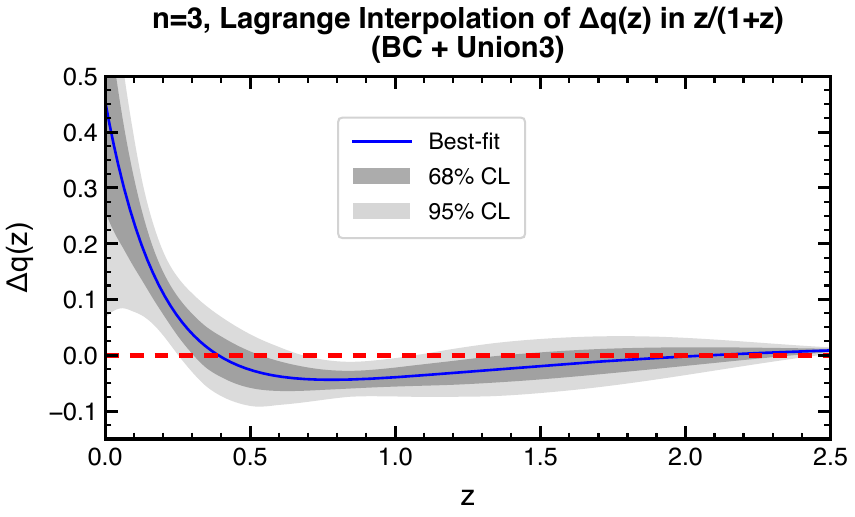}
\end{minipage}

\vspace{1ex}
\begin{minipage}{0.32\textwidth}
  \centering
  \includegraphics[width=\textwidth]{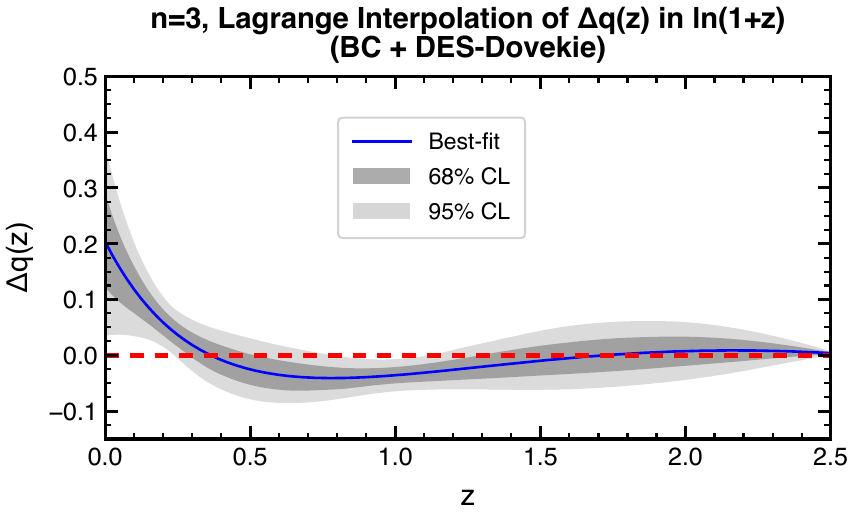}
\end{minipage}\hfill
\begin{minipage}{0.32\textwidth}
  \centering
  \includegraphics[width=\textwidth]{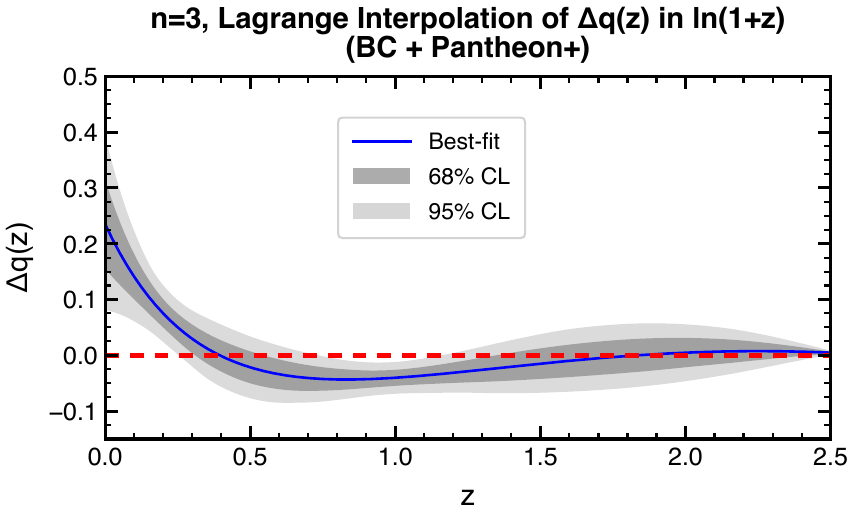}
\end{minipage}\hfill
\begin{minipage}{0.32\textwidth}
  \centering
  \includegraphics[width=\textwidth]{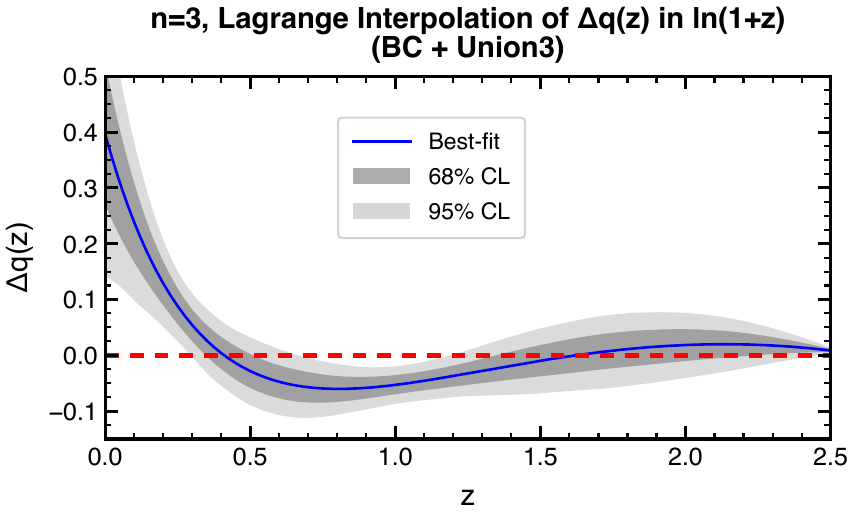}
\end{minipage}

\caption{Reconstruction of the deviation of the deceleration parameter from the best-fit $\Lambda$CDM prediction, $\Delta q(z)\equiv q_{\rm rec}(z)-q_{\Lambda{\rm CDM}}^{\rm best\text{-}fit}(z)$, using third-order Lagrange interpolation. Rows (top to bottom) correspond to adopting $z$, $y\equiv z/(1+z)$, and $\zeta\equiv\ln(1+z)$ as the interpolation variables, respectively. Columns (left to right) correspond to the DES-Dovekie, Pantheon+, and Union3 SNe Ia compilations combined with DESI BAO and CMB data. The solid blue curve shows the best-fit reconstruction of $\Delta q(z)$, with the shaded regions denoting the $68\%$ (darker gray) and $95\%$ (lighter gray) conﬁdence levels (CLs).}
\label{fig:cub_dqz_interp}
\end{figure*}

\begin{figure*}[t]
\centering
\begin{minipage}{0.32\textwidth}
  \centering
  \includegraphics[width=\textwidth]{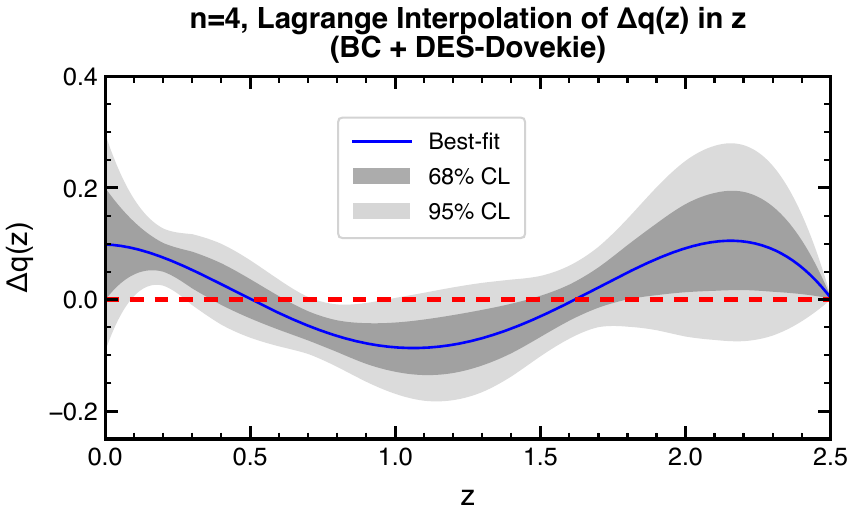}
\end{minipage}\hfill
\begin{minipage}{0.32\textwidth}
  \centering
  \includegraphics[width=\textwidth]{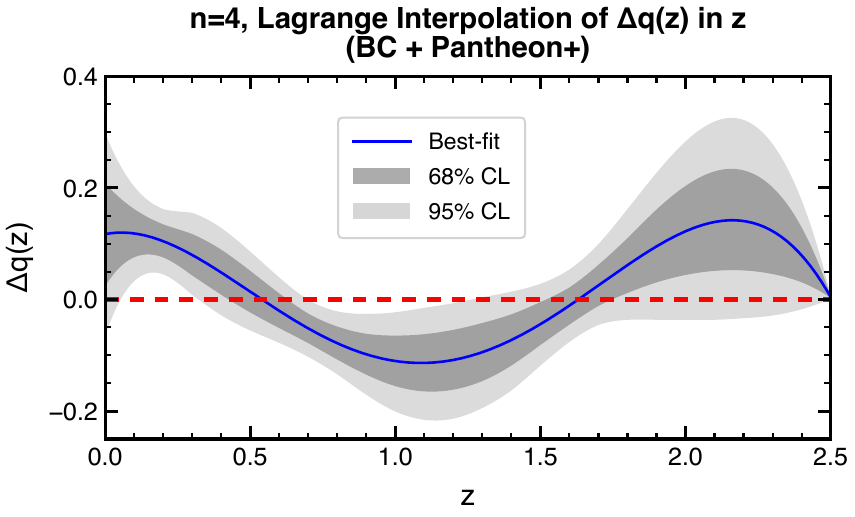}
\end{minipage}\hfill
\begin{minipage}{0.32\textwidth}
  \centering
  \includegraphics[width=\textwidth]{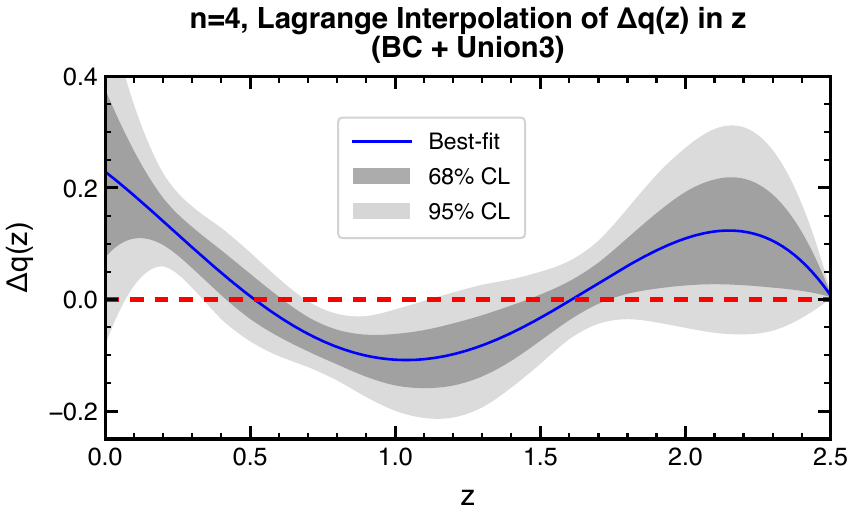}
\end{minipage}

\vspace{1ex} 
\begin{minipage}{0.32\textwidth}
  \centering
  \includegraphics[width=\textwidth]{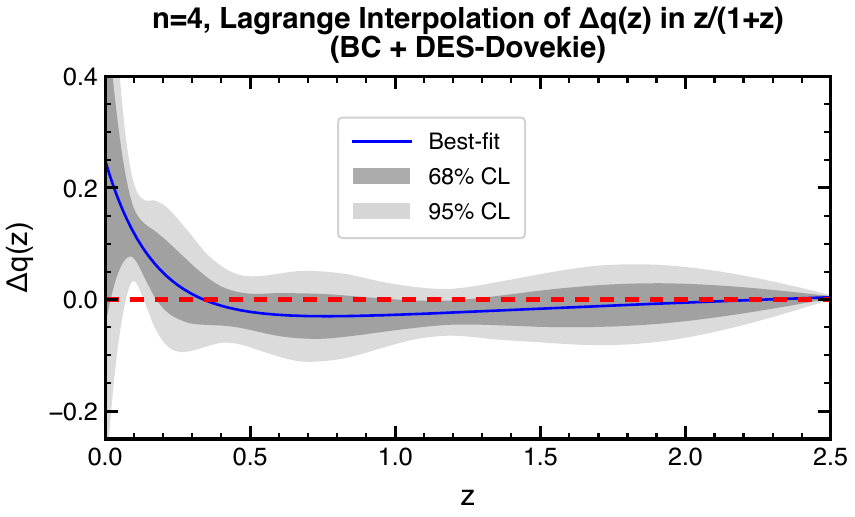}
\end{minipage}\hfill
\begin{minipage}{0.32\textwidth}
  \centering
  \includegraphics[width=\textwidth]{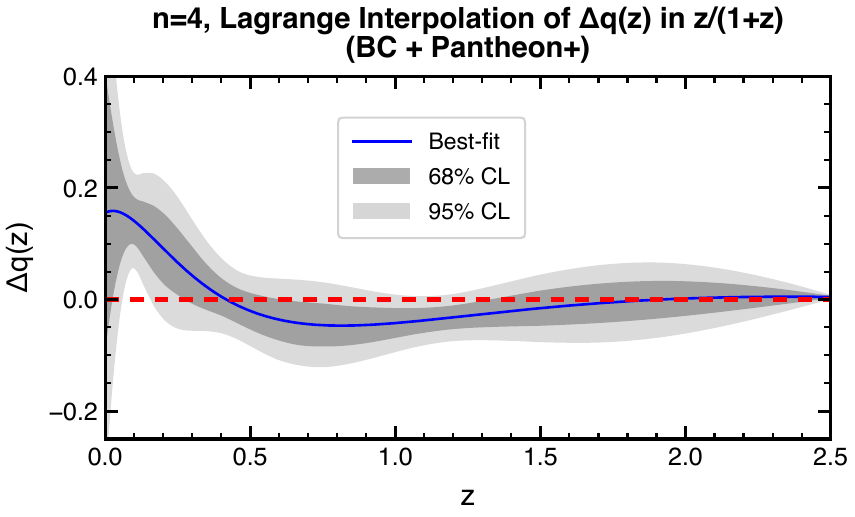}
\end{minipage}\hfill
\begin{minipage}{0.32\textwidth}
  \centering
  \includegraphics[width=\textwidth]{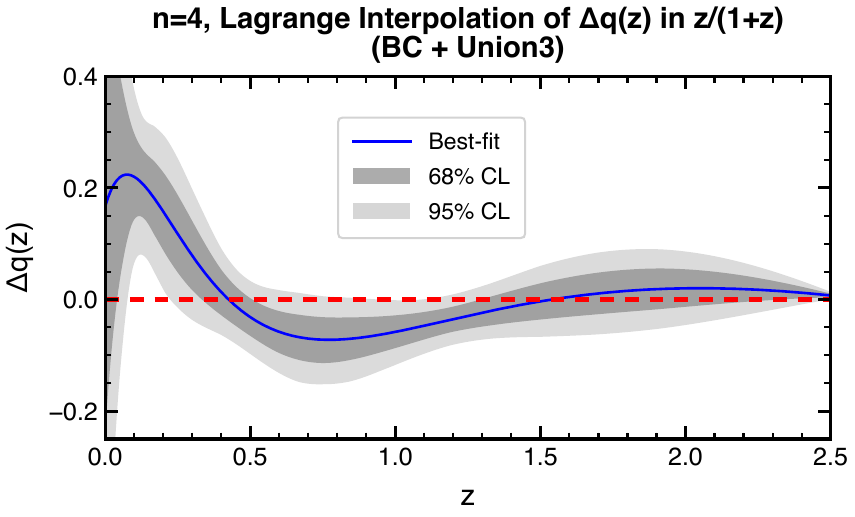}
\end{minipage}

\vspace{1ex}
\begin{minipage}{0.32\textwidth}
  \centering
  \includegraphics[width=\textwidth]{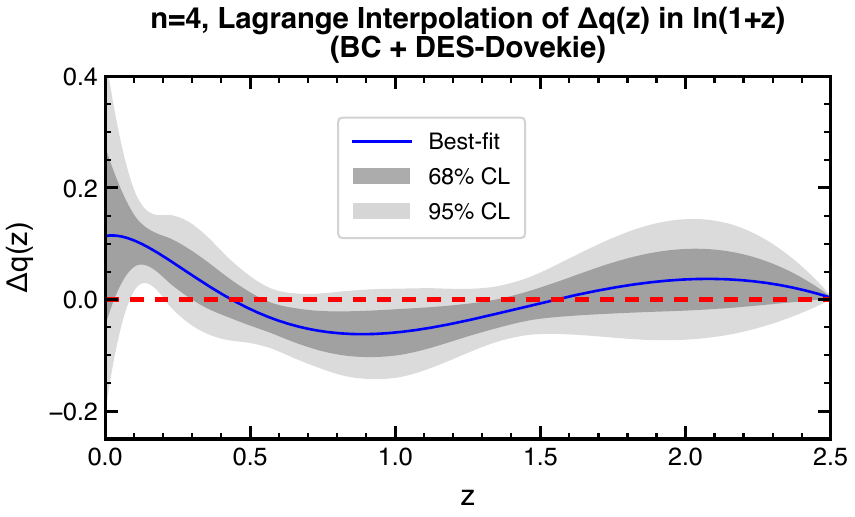}
\end{minipage}\hfill
\begin{minipage}{0.32\textwidth}
  \centering
  \includegraphics[width=\textwidth]{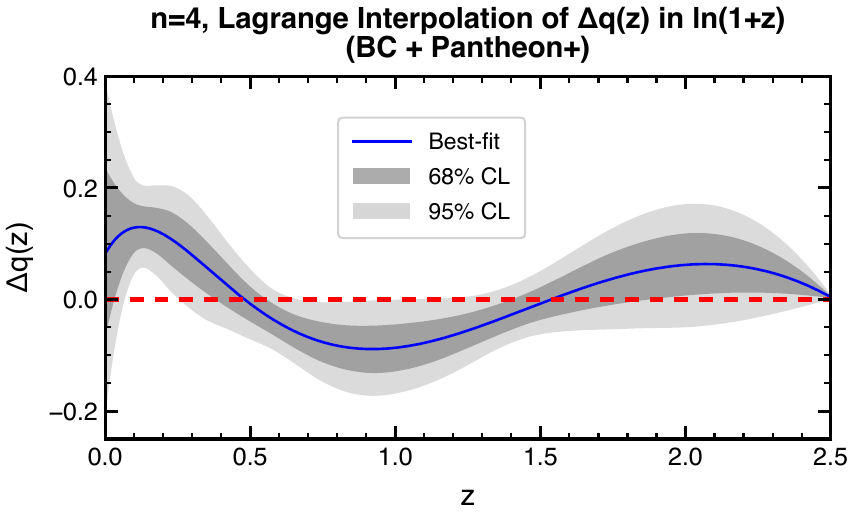}
\end{minipage}\hfill
\begin{minipage}{0.32\textwidth}
  \centering
  \includegraphics[width=\textwidth]{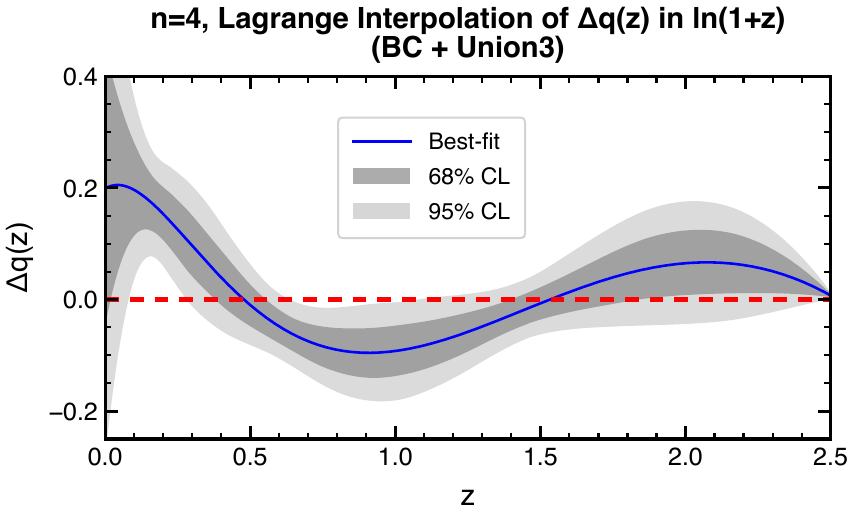}
\end{minipage}
\caption{Reconstruction of the deviation of the deceleration parameter from the best-fit $\Lambda$CDM prediction, $\Delta q(z)\equiv q_{\rm rec}(z)-q_{\Lambda{\rm CDM}}^{\rm best\text{-}fit}(z)$, using fourth-order Lagrange interpolation. Here, $q_{\rm rec}(z)$ is given by Eq.~\eqref{qzdqz}, while $q_{\Lambda{\rm CDM}}^{\rm best\text{-}fit}(z)$ is obtained from Eq.~\eqref{qLCDM} by setting $\Omega_{m,0}$ to its best-fit value. Rows (top to bottom) correspond to adopting $z$, $y\equiv z/(1+z)$, and $\zeta\equiv \ln(1+z)$ as the interpolation variables, respectively. Columns (left to right) show the results for the DES-Dovekie, Pantheon+, and Union3 SNe Ia compilations combined with DESI BAO and CMB data. The solid blue curve shows the best-fit reconstruction of $\Delta q(z)$ with the shaded regions denoting the $68\%$ (darker gray) and $95\%$ (lighter gray) CLs.}
\label{fig:quart_dqz_interp}
\end{figure*}

\begin{figure*}[t]
\centering
\begin{minipage}{0.32\textwidth}
  \centering
  \includegraphics[width=\textwidth]{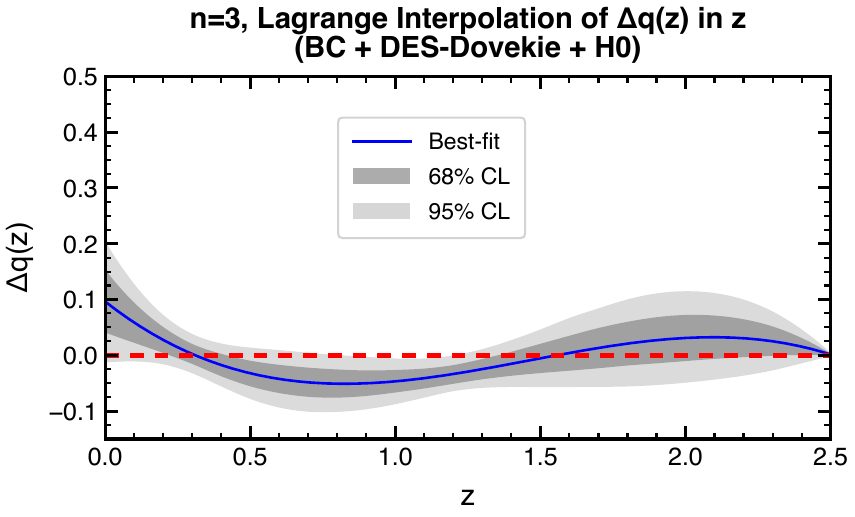}
\end{minipage}\hfill
\begin{minipage}{0.32\textwidth}
  \centering
  \includegraphics[width=\textwidth]{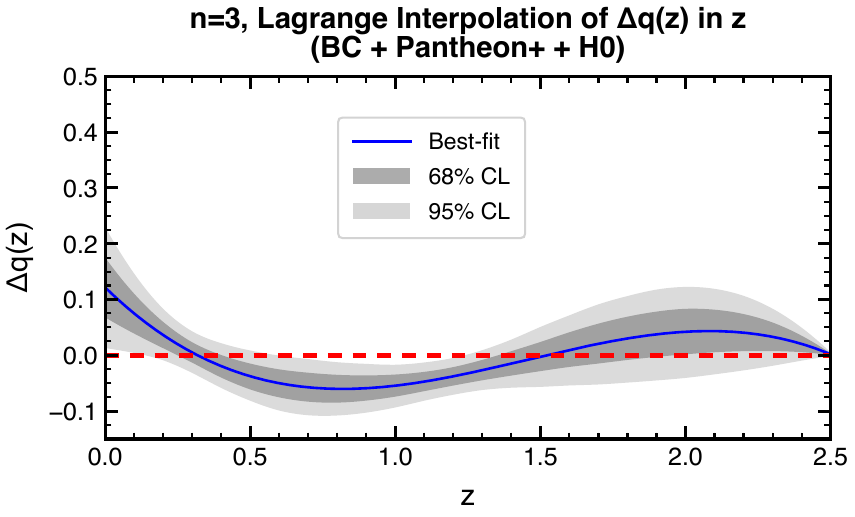}
\end{minipage}\hfill
\begin{minipage}{0.32\textwidth}
  \centering
  \includegraphics[width=\textwidth]{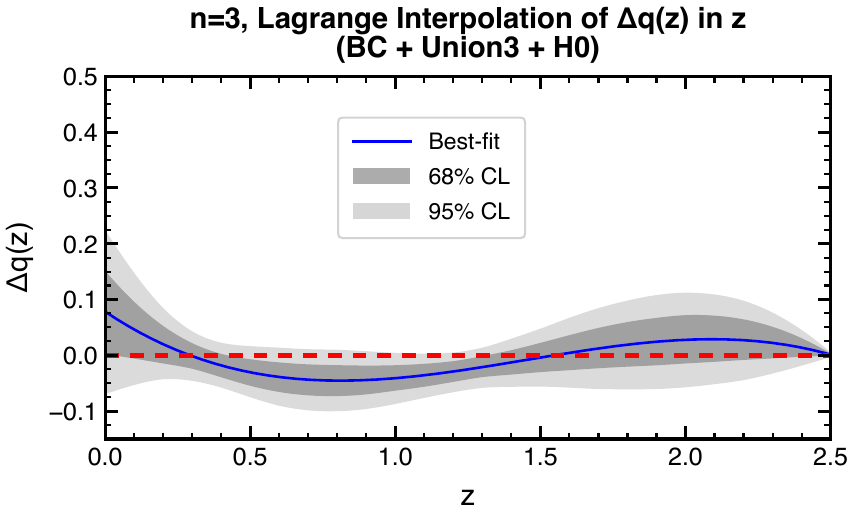}
\end{minipage}

\vspace{1ex}
\begin{minipage}{0.32\textwidth}
  \centering
  \includegraphics[width=\textwidth]{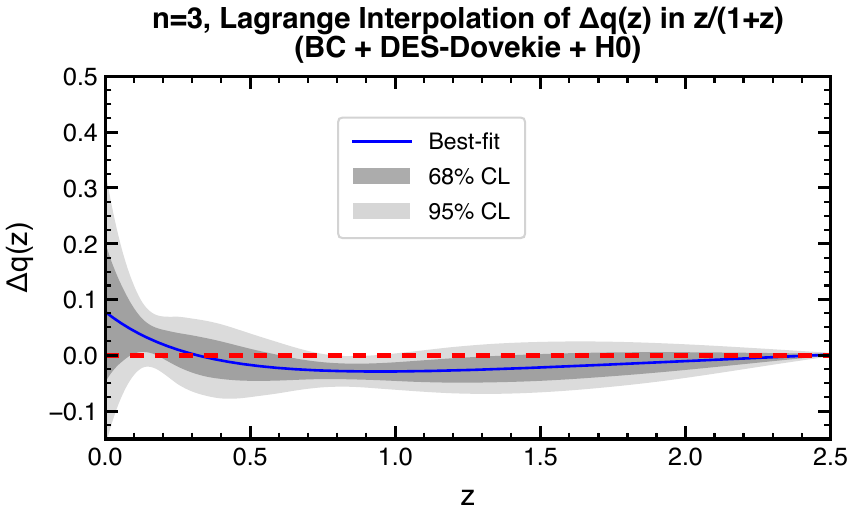}
\end{minipage}\hfill
\begin{minipage}{0.32\textwidth}
  \centering
  \includegraphics[width=\textwidth]{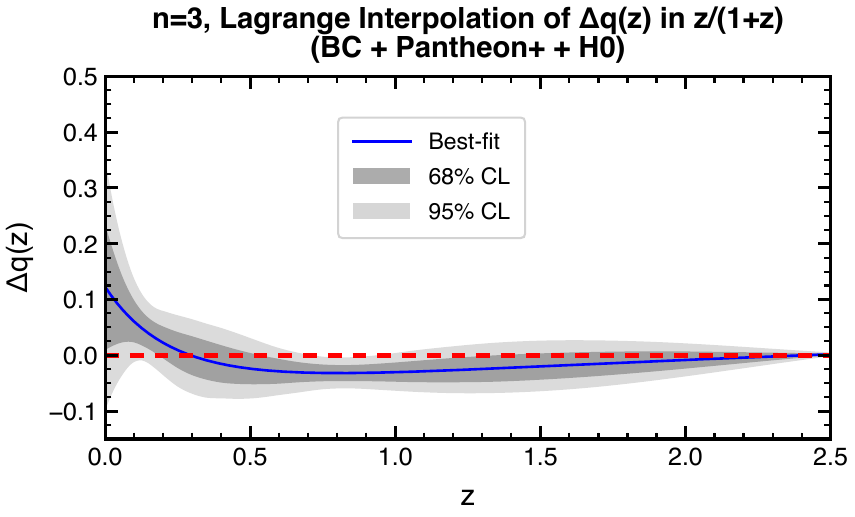}
\end{minipage}\hfill
\begin{minipage}{0.32\textwidth}
  \centering
  \includegraphics[width=\textwidth]{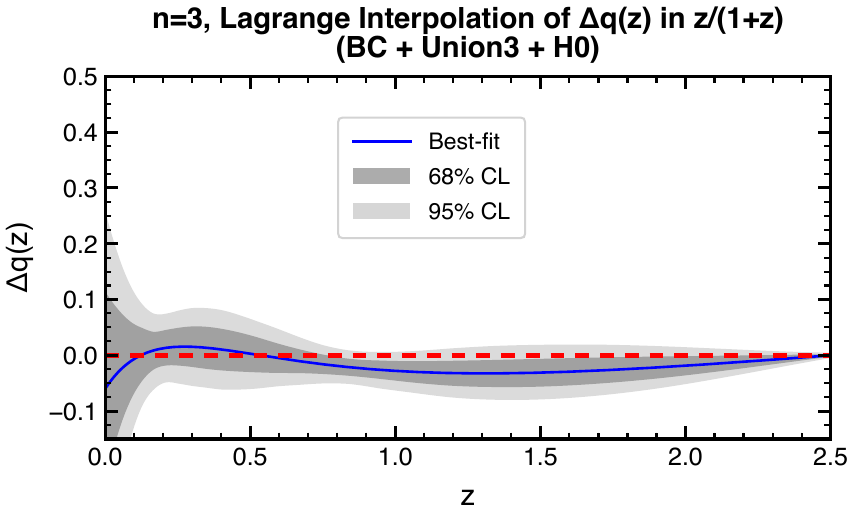}
\end{minipage}

\vspace{1ex}
\begin{minipage}{0.32\textwidth}
  \centering
  \includegraphics[width=\textwidth]{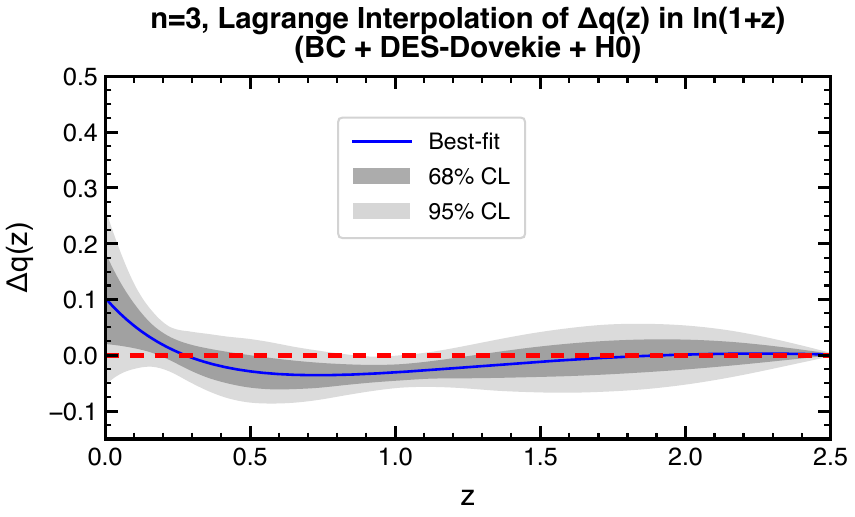}
\end{minipage}\hfill
\begin{minipage}{0.32\textwidth}
  \centering
  \includegraphics[width=\textwidth]{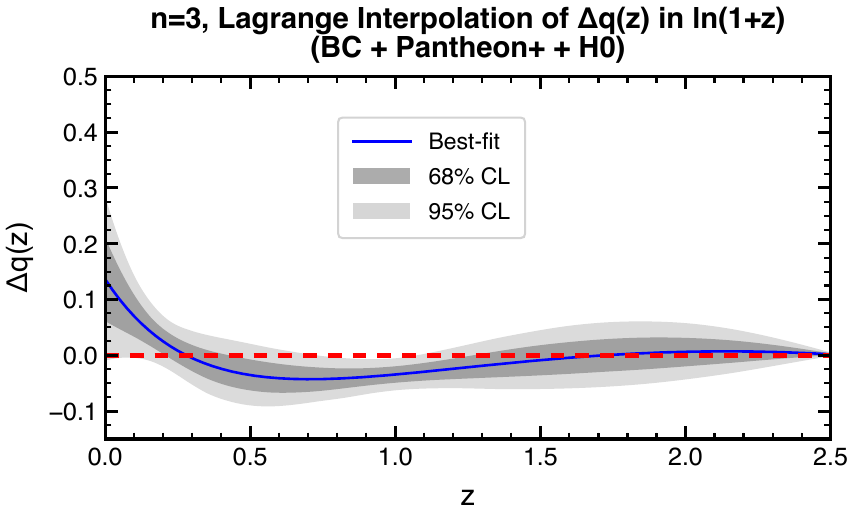}
\end{minipage}\hfill
\begin{minipage}{0.32\textwidth}
  \centering
  \includegraphics[width=\textwidth]{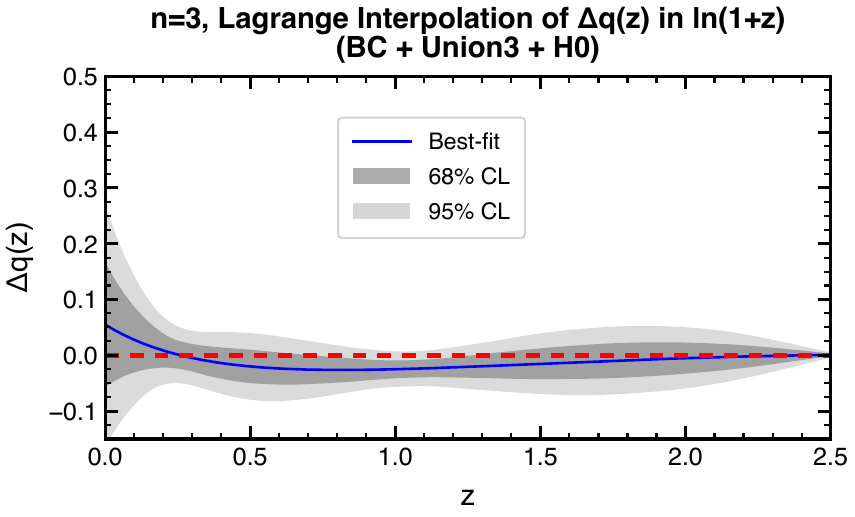}
\end{minipage}
\caption{Same as Fig.~\ref{fig:cub_dqz_interp}, but with an $H_0$ prior from SH0ES included in each data combination.}
\label{fig:cub_dqz_interp_H0}
\end{figure*}

\begin{figure*}[t]
\centering
\begin{minipage}{0.32\textwidth}
  \centering
  \includegraphics[width=\textwidth]{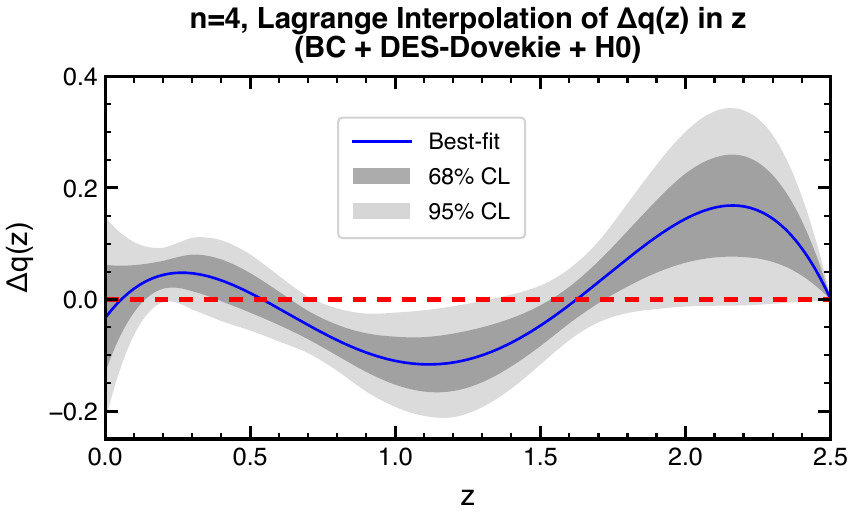}
\end{minipage}\hfill
\begin{minipage}{0.32\textwidth}
  \centering
  \includegraphics[width=\textwidth]{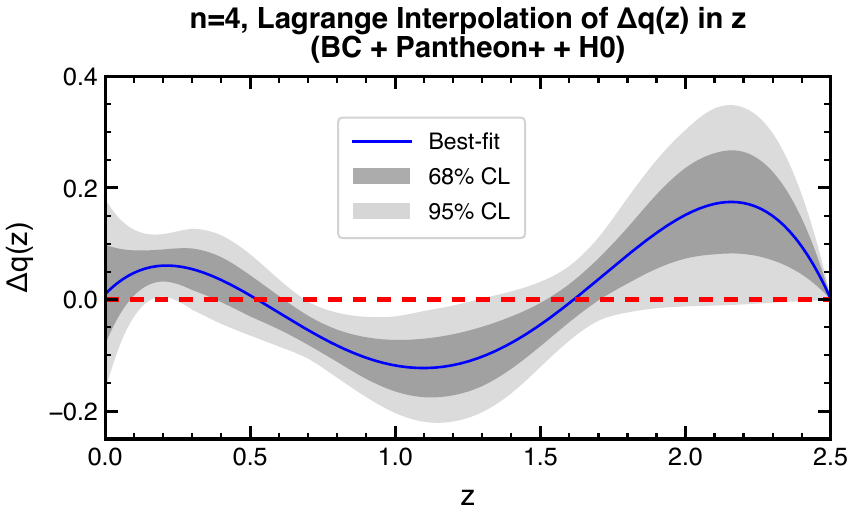}
\end{minipage}\hfill
\begin{minipage}{0.32\textwidth}
  \centering
  \includegraphics[width=\textwidth]{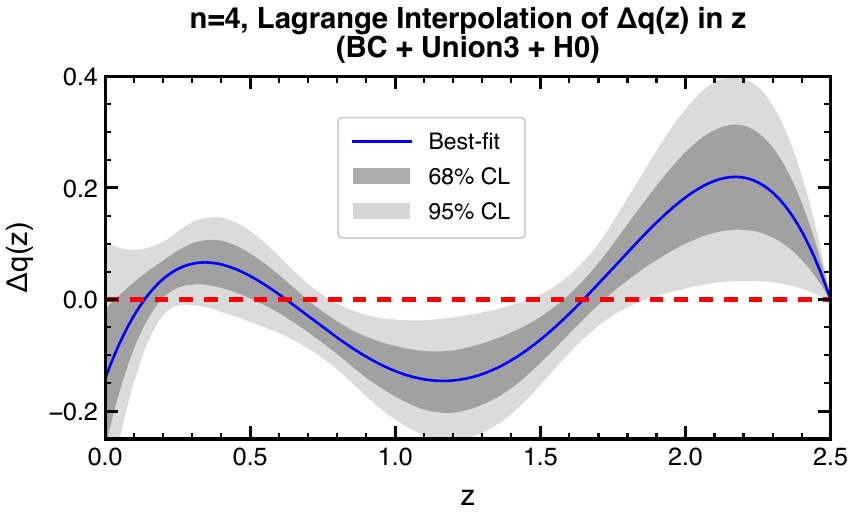}
\end{minipage}

\vspace{1ex}
\begin{minipage}{0.32\textwidth}
  \centering
  \includegraphics[width=\textwidth]{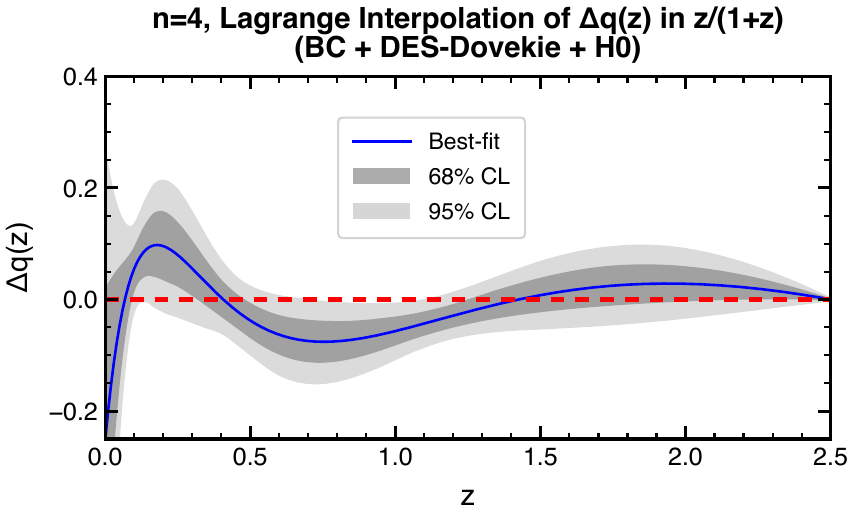}
\end{minipage}\hfill
\begin{minipage}{0.32\textwidth}
  \centering
  \includegraphics[width=\textwidth]{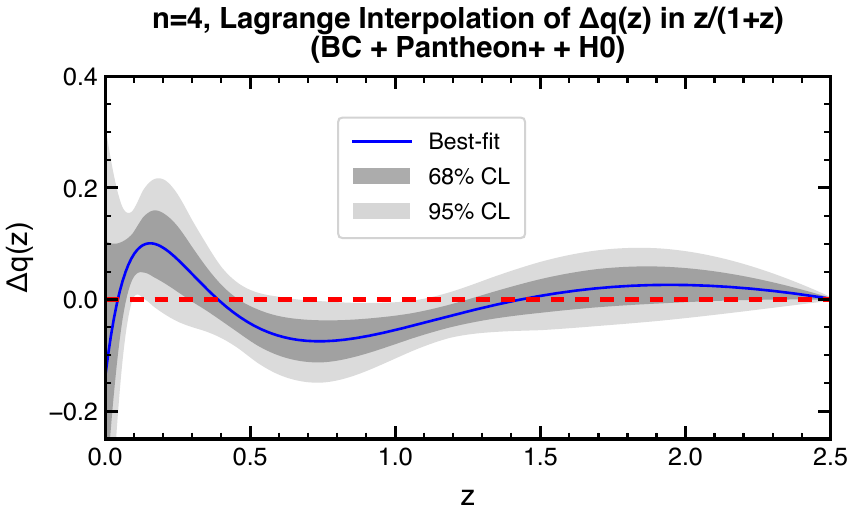}
\end{minipage}\hfill
\begin{minipage}{0.32\textwidth}
  \centering
  \includegraphics[width=\textwidth]{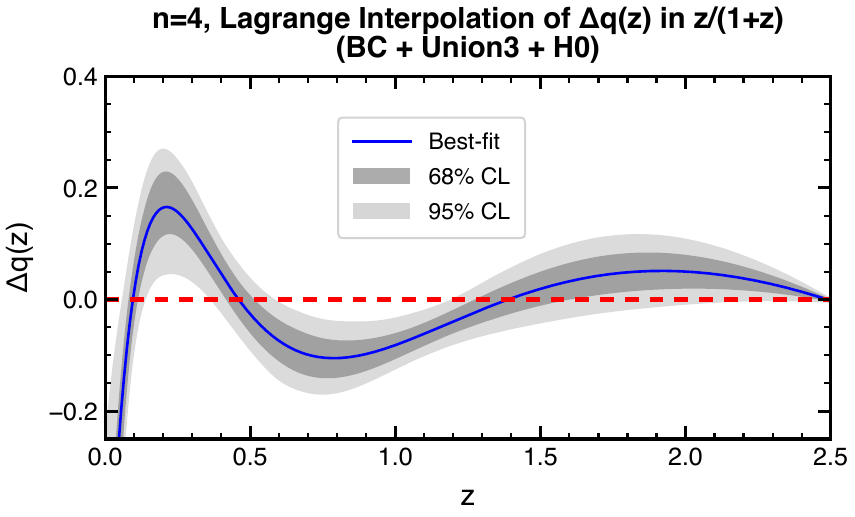}
\end{minipage}

\vspace{1ex}
\begin{minipage}{0.32\textwidth}
  \centering
  \includegraphics[width=\textwidth]{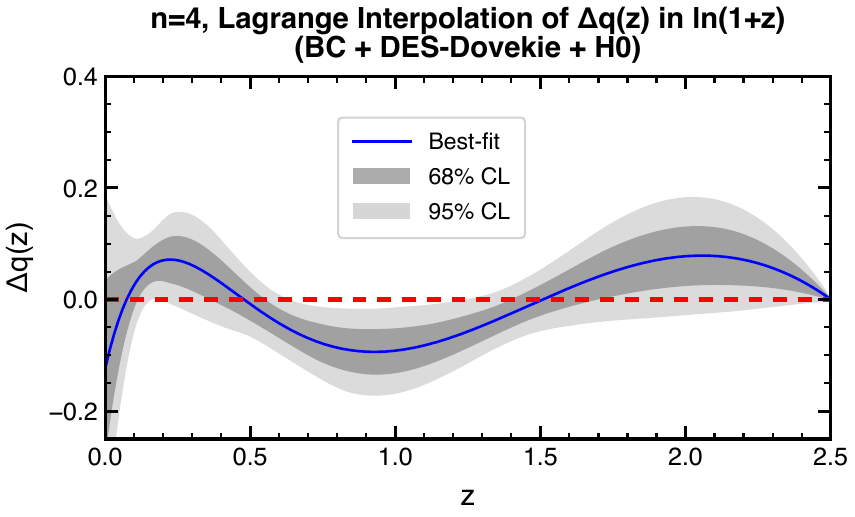}
\end{minipage}\hfill
\begin{minipage}{0.32\textwidth}
  \centering
  \includegraphics[width=\textwidth]{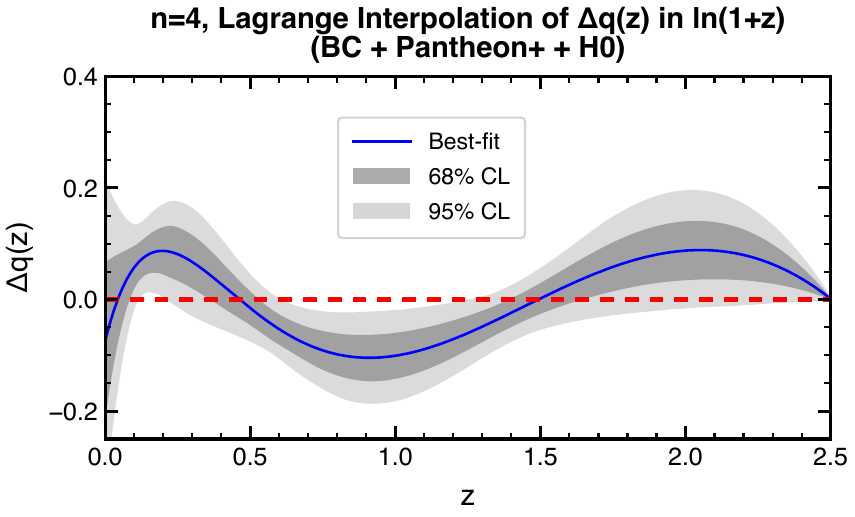}
\end{minipage}\hfill
\begin{minipage}{0.32\textwidth}
  \centering
  \includegraphics[width=\textwidth]{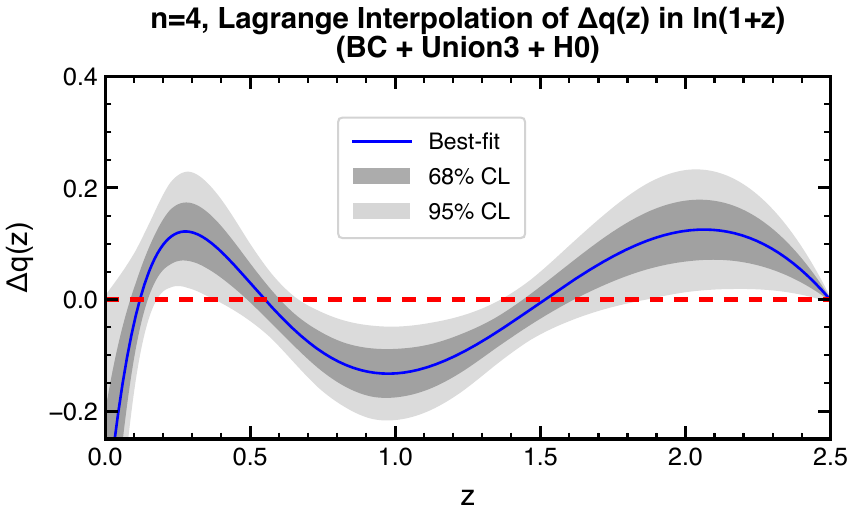}
\end{minipage}
\caption{Same as Fig.~\ref{fig:quart_dqz_interp}, but with an $H_0$ prior from SH0ES included in each data combination.}
\label{fig:quart_dqz_interp_H0}
\end{figure*}

We reconstruct $q(z)$ to probe the redshift evolution of the deceleration parameter, using the joint datasets from DESI BAO and CMB (hereafter referred to as BC), along with DES-Dovekie/Pantheon+/Union3. The constraint results, including the mean values and $68\%$ CL for the cosmological and derived parameters, obtained without and with the SH0ES $H_0$ prior, are summarized in Tables~\ref{tab:parameters} and \ref{tab:parameters_H0}, respectively. The reconstructed $\Delta q(z)$ curves, where $\Delta q(z)\equiv q_{\rm rec}(z)-q_{\Lambda{\rm CDM}}^{\rm best\text{-}fit}(z)$, obtained from cubic and quartic Lagrange interpolations, respectively, with/without the $H_0$ prior, are illustrated in Figs.~\ref{fig:cub_dqz_interp}, \ref{fig:quart_dqz_interp}, \ref{fig:cub_dqz_interp_H0}, and \ref{fig:quart_dqz_interp_H0}. Additionally, we use Bayesian evidence and frequentist significance to assess the performance of these Lagrange interpolation reconstructions. The resulting difference in chi-square, $\Delta\chi^2_{\rm min} \equiv \chi^2_{\rm min} - \chi^2_{\rm min}(\Lambda{\rm CDM})$, the corresponding significance $\sigma$ obtained from $\Delta\chi^2_{\rm min}$ via Eq.~(\ref{eq:frequentist}) based on Wilks' theorem~\cite{Wilks1938}, and the logarithmic Bayes factors $\ln B_{ij}$, with the $\Lambda$CDM model as the reference, are summarized in Table~\ref{tab:lnB_withH0}.

We first discuss the constraints on cosmological parameters: $H_0$, $\Omega_{m,0}h^2$ and $\Omega_{b,0}h^2$. As shown in Tables~\ref{tab:parameters} and \ref{tab:parameters_H0}, the Lagrange polynomial constraints from BC favor lower central values of $H_0$ than the Planck 2018 estimate ($H_0 = 67.4\pm0.5$\,$\mathrm{km\,s^{-1}\,Mpc^{-1}}$~\cite{PlanckCollaboration2020A&A}), with significances ranging from $0.5\sigma$ to $3.2\sigma$. The lower central values of $H_0$ inferred from the BC-only reconstructions are qualitatively consistent with the downward shift in $H_0$ found in the DESI DR2 analysis of DESI+CMB when extending $\Lambda$CDM to $w_0w_a$CDM. This behavior may be related to the small discrepancy between the DESI DR2 BAO distance measurements and the Planck predictions, as shown in Fig.~6 of Ref.~\cite{AbdulKarim2025PhRvD}. After incorporating SNe Ia data, the Lagrange polynomial yields values of $H_0$ that are all consistent with the Planck 2018 estimate, but exhibit a tension of $\geq4.3\sigma$ with the SH0ES determination~\cite{Riess2022ApJL}. Across all data combinations, the reconstructed constraints on both $\Omega_{m,0}h^2$ and $\Omega_{b,0}h^2$ remain consistent with the corresponding $\Lambda$CDM results, which reflects that they are  tightly determined  by the CMB distance priors. Among the SNe Ia samples without the $H_0$ prior, the BC+Union3 combination exhibits the largest departures from $\Lambda$CDM, while BC+DES-Dovekie is the closest to it. Notably, imposing the SH0ES $H_0$ prior on the BC+SNe Ia combinations has negligible effects on the $\Lambda$CDM constraints. This is because $\Lambda$CDM has only three free parameters, $H_0$, $\Omega_{m,0}h^2$, and $\Omega_{b,0}h^2$, which are already primarily and tightly constrained by the BC+SNe Ia data.

Three derived parameters $q_0$, $z_t$, and $q_{\rm pivot}\equiv q(z=0.7)$, characterizing respectively the present deceleration parameter, the redshift of transition from decelerating to accelerating cosmic expansion, and the expansion state at intermediate redshift, are also discussed in our analysis. We choose $z=0.7$ as the pivot redshift because the reconstructed $q(z)$ has a relatively small uncertainty at this redshift and is only weakly affected by the cutoff condition at $z_c$. The constraints on them  are summarized in Table~\ref{tab:parameters}. For the $\Lambda$CDM model, BC yield $q_0=-0.55\pm0.01$, providing  strong evidence for currently accelerating cosmic expansion. However, this evidence diminishes to less than $1\sigma$  in the Lagrange interpolation models. Notably, in cubic-$\zeta$ interpolation, BC even favor a currently decelerating expansion. When analyzing the SNe Ia data, such as Pantheon+,  in isolation  to constrain $q_0$, we find from Table~\ref{tab:parameters} that in $\Lambda$CDM $q_0$ is less than zero at a very high confidence level. Additionally, all cubic interpolations and the quartic-$z$ interpolation also favor $q_0<0$, whereas the other interpolation models show no clear preference. Combining BC with SNe Ia significantly tightens the constraint on $q_0$. For all models except quartic-$y$, both BC+Pantheon+ and BC+DES-Dovekie support $q_0<0$ at over $2\sigma$. Only in the cubic-$y$ and quartic-$y$ interpolations is $q_0=0$ allowed within $1\sigma$ by BC+Union3. For the BC and BC+SNe Ia data combinations, the Lagrange-polynomial reconstructions indicate a weaker present-day acceleration than that of $\Lambda$CDM (i.e., a less negative $q_0$) in most cases. For the BC+SNe Ia data combinations, for example, the cubic-$z$ constraints on $q_0$ deviate from $\Lambda$CDM at more than $2\sigma$, whereas the quartic-$z$ reconstruction, which has larger uncertainties, remains statistically consistent with $\Lambda$CDM. When the SH0ES prior is imposed, the posteriors of $q_0$ shift toward stronger acceleration (more negative $q_0$), and the central estimates become more negative than the $\Lambda$CDM value for the cubic-$y$ and all quartic cases with BC+Union3+$H_0$, as illustrated in Table \ref{tab:parameters_H0}.

Regarding the transition redshift $z_t$, all reconstructions favor an earlier transition from matter domination to dark energy domination than predicted by $\Lambda$CDM across all datasets. This finding is consistent with what was obtained in Ref.~\cite{Lodha2025PhRvD}, where DESI Collaboration derived $q(z)$ from BAO, CMB, and three SNe Ia datasets within the $w_{0}w_{a}$CDM framework. In the $\Lambda$CDM model, $q_{\rm pivot}$ is slightly positive for all data combinations, whereas all reconstructions yield negative central values, with the BC-only case giving the most negative values for $q_{\rm pivot}$. This result also indicates that reconstructions  favor the Universe entering the accelerating phase slightly earlier than predicted by the $\Lambda$CDM model. After imposing the SH0ES prior on the BC+SNe Ia combinations, $q_{\rm pivot}$ generally decreases for both $\Lambda$CDM and the reconstructed models, with the exception of the Union3 combinations in the cubic reconstructions and the quartic-$z$ reconstruction. Additionally, the posterior uncertainties for $q_0$, $q_{\rm pivot}$, and $z_t$ are generally smaller for the cubic reconstructions compared to the quartic ones.

\begin{figure}[t]
\centering
\begin{minipage}{0.24\textwidth}
  \centering
  \includegraphics[width=\textwidth]{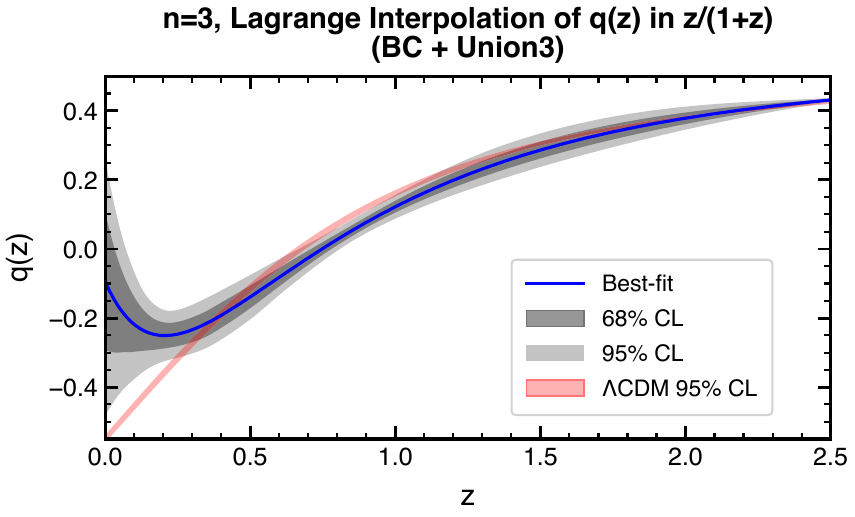}
\end{minipage}\hfill
\begin{minipage}{0.24\textwidth}
  \centering
  \includegraphics[width=\textwidth]{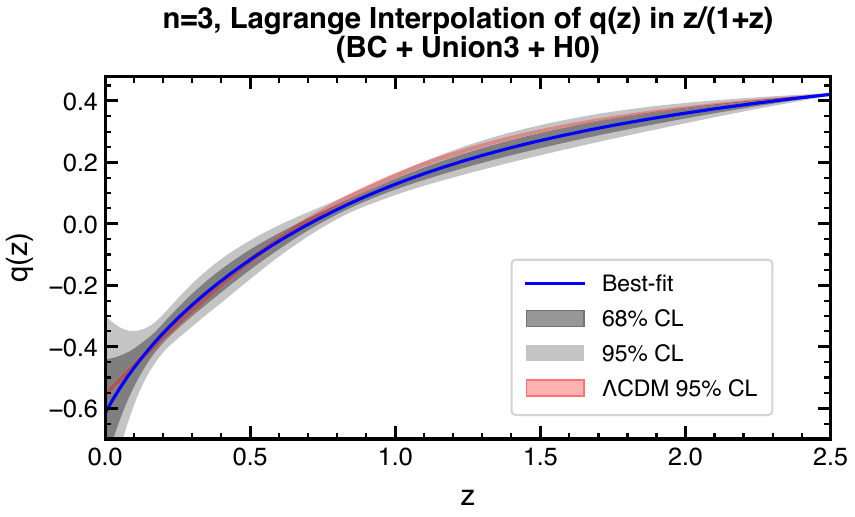}
\end{minipage}\hfill

\vspace{1ex}
\begin{minipage}{0.24\textwidth}
  \centering
  \includegraphics[width=\textwidth]{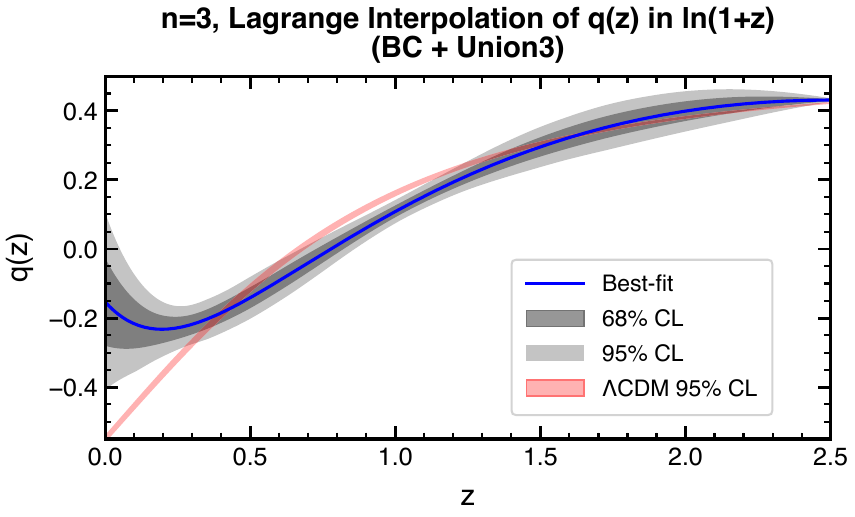}
\end{minipage}\hfill
\begin{minipage}{0.24\textwidth}
  \centering
  \includegraphics[width=\textwidth]{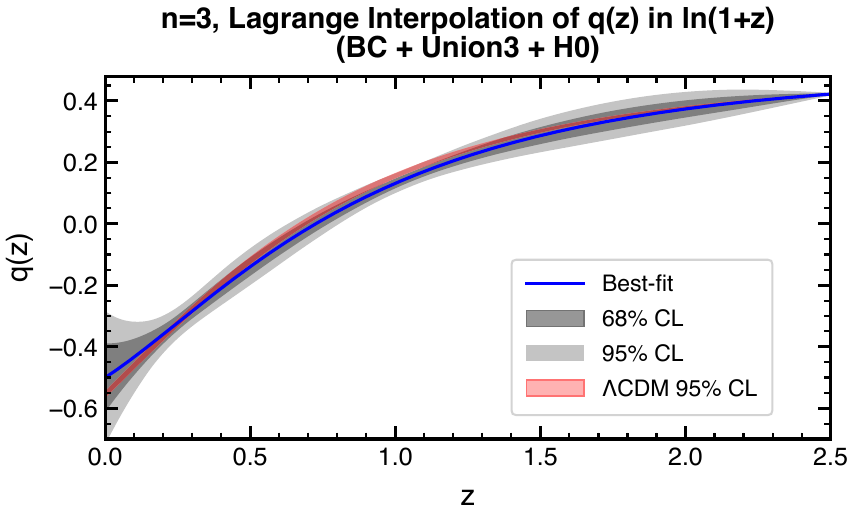}
\end{minipage}\hfill
\caption{The deceleration parameter $q(z)$ reconstructed using third-order Lagrange interpolation. The top row adopts $y\equiv z/(1+z)$, while the bottom row adopts $\zeta\equiv\ln(1+z)$ as the interpolation variables, respectively. The left column corresponds to BC+Union3, while the right column corresponds to BC+Union3+$H_0$. Blue solid curves show the best-fit reconstruction of $q(z)$, with shaded regions showing the 68\% (darker gray) and 95\% (lighter gray) CLs. The red shaded regions denote the 95\% CLs of the $\Lambda$CDM model.}
\label{fig:cubic_qz_interp}
\end{figure}

Next, we focus on the evolution of the deviation of the reconstructed deceleration parameter from the best-fit $\Lambda$CDM prediction, $\Delta q(z)\equiv q_{\rm rec}(z)-q_{\Lambda{\rm CDM}}^{\rm best\text{-}fit}(z)$, using the BC+DES-Dovekie/Pantheon+/Union3 combinations, where $q_{\rm rec}(z)$ is obtained from the MCMC chains using Eq.~\eqref{qzdqz}. The results are presented in Figs.~\ref{fig:cub_dqz_interp} and \ref{fig:quart_dqz_interp}. The reconstructed $\Delta q(z)$ curves (solid blue) lie significantly above the $\Lambda$CDM prediction (red dashed) at $z\lesssim0.3$, with $\Lambda$CDM falling outside the $95\%$ credible region for the cubic reconstructions and the $68\%$ credible region for the majority of quartic reconstructions. This indicates a clear deviation from $\Lambda$CDM at low redshifts. Moreover, the reconstructed $\Delta q(z)$ alternates between lying above and below the $\Lambda$CDM prediction. Specifically, it is positive in the ranges $0 \lesssim z \lesssim 0.5$ and $1.5 \lesssim z \lesssim 2.5$, while it is negative in the range $0.5 \lesssim z \lesssim 1.5$. This results in an overall oscillatory pattern around the $\Lambda$CDM model throughout the evolution, consistent with the findings of DESI Collaboration~\cite{Lodha2025PhRvD}. Comparing three redshift expansion forms, we find that while the $z$-expansion yields smaller uncertainty bands than the $y$- and $\zeta$-expansions at low redshifts, it develops significantly broader uncertainty bands at high redshifts. The cosmic expansion at high redshifts is expected to approach the $\Lambda$CDM prediction. However, Figs.~\ref{fig:cub_dqz_interp} and \ref{fig:quart_dqz_interp} show that the reconstructed evolution using the $z$-expansion deviates substantially from the $\Lambda$CDM model at high redshifts. This deviation indicates that using $z$ as the expansion variable is less consistent with our expectations than employing $y$ or $\zeta$. Furthermore, we examine whether cosmic acceleration has already reached its maximum. Among all the cases considered, only the BC+Union3 cubic-$y$ and cubic-$\zeta$ reconstructions indicate that cosmic acceleration has already reached its maximum at low $z$ and is beginning to slow down; these results are shown in the left column of Fig.~\ref{fig:cubic_qz_interp}. This finding is similar to that obtained by Wang et al.~\cite{Wang2025EPJC}, who provided evidence for a late-time slowing of cosmic acceleration in the BC+Union3/DES5YR samples within a flat $w_0w_a$CDM framework. 

Now, we assess the impact of imposing the local $H_0$ prior and present the results in Figs.~\ref{fig:cub_dqz_interp_H0} and \ref{fig:quart_dqz_interp_H0}. Comparing Fig.~\ref{fig:cub_dqz_interp_H0} with Fig.~\ref{fig:cub_dqz_interp}, we find that, with the local $H_0$ prior, all cubic best-fit $\Delta q(z)$ curves shift toward zero, which corresponds to the $\Lambda$CDM prediction, particularly at low redshift. 
Consequently, the inclusion of the $H_0$ prior reduces the low-redshift amplitude of the best-fit $\Delta q(z)$ curve in the cubic reconstructions. Overall, imposing the SH0ES $H_0$ prior weakens the apparent signatures of dynamical dark energy in the cubic reconstructions, including a reduced magnitude of the present-day deviation (from about $0.2$--$0.4$ to about $0.1$) and a late-time decline in the cosmic acceleration. This can also be seen from Fig.~\ref{fig:cubic_qz_interp}, which shows that, once the $H_0$ prior is included, the low-redshift bump disappears in both the cubic-$y$ and cubic-$\zeta$ reconstructions, and no peak in the acceleration is observed. This weakening suggests a potential tension between the SH0ES $H_0$ measurement and the dynamical features favored by the DESI DR2 BAO data. These findings are consistent with the results in Ref.~\cite{Pang2025SCPMA}, which showed that adding the SH0ES prior weakens the preference for dynamical dark energy, highlighting a tension between the SH0ES $H_0$ value and the phantom-to-quintessence transition in dark energy favored by the DESI DR2 BAO data. However, in the quartic case, a comparison of Figs.~\ref{fig:quart_dqz_interp_H0} and~\ref{fig:quart_dqz_interp} reveals that after imposing the SH0ES $H_0$ prior, the best-fit $\Delta q(z)$ once again changes sign at $z\lesssim 0.15$ and becomes negative. This indicates  stronger cosmic acceleration than predicted by the $\Lambda$CDM model. Furthermore, regardless of whether the $H_0$ prior is included, the reconstructions using $y\equiv z/(1+z)$ and $\zeta\equiv\ln(1+z)$ exhibit  smaller $1\sigma$ uncertainties compared to the standard $z$-based form at high redshifts. This advantage arises from their effectiveness in addressing the highly nonuniform redshift distribution of SNe Ia and BAO data, which is dense at low redshifts and sparse at high redshifts.

\begin{table*}[t]
\caption{Model comparison in terms of $\Delta\chi^2_{\rm min}\equiv \chi^2_{\rm min}-\chi^2_{\rm min}(\Lambda{\rm CDM})$, the corresponding significance $\sigma$ obtained from $\Delta\chi^2_{\rm min}$ via Eq.~(\ref{eq:frequentist}) based on Wilks' theorem~\cite{Wilks1938}, and the log Bayes factor $\ln B_{ij}$, where $j$ denotes the $\Lambda$CDM reference model. Negative values of $\ln B_{ij}$ and positive values of $\Delta\chi^2_{\rm min}$ indicate preference for $\Lambda$CDM. Here, BC denotes the combination of DESI BAO and CMB data.}
\label{tab:lnB_withH0}
\centering
\scriptsize
\setlength{\tabcolsep}{2pt}
\renewcommand{\arraystretch}{1.2}

\begin{ruledtabular}
\begin{tabular}{l c c c c c c c c c c c c c c c c c c}
\multirow{2}{*}{\textbf{Dataset}}
& \multicolumn{3}{c}{\textbf{Cubic-$z$}}
& \multicolumn{3}{c}{\textbf{Cubic-$y$}}
& \multicolumn{3}{c}{\textbf{Cubic-$\zeta$}}
& \multicolumn{3}{c}{\textbf{Quartic-$z$}}
& \multicolumn{3}{c}{\textbf{Quartic-$y$}}
& \multicolumn{3}{c}{\textbf{Quartic-$\zeta$}} \\
& $\Delta\chi^2_{\rm min}$ & $\sigma$ & $\ln B_{ij}$
& $\Delta\chi^2_{\rm min}$ & $\sigma$ & $\ln B_{ij}$
& $\Delta\chi^2_{\rm min}$ & $\sigma$ & $\ln B_{ij}$
& $\Delta\chi^2_{\rm min}$ & $\sigma$ & $\ln B_{ij}$
& $\Delta\chi^2_{\rm min}$ & $\sigma$ & $\ln B_{ij}$
& $\Delta\chi^2_{\rm min}$ & $\sigma$ & $\ln B_{ij}$ \\
\hline
BC+DES-Dovekie & $-10.5$ & $2.4$ & $-5.0$ & $-9.7$ & $2.3$ & $-5.2$ & $-9.9$ & $2.3$ & $-5.1$ & $-11.2$ & $2.3$ & $-6.5$ & $-9.7$ & $2.0$ & $-6.1$ & $-10.3$ & $2.1$ & $-6.5$ \\
BC+DES-Dovekie+$H_0$ & $-5.9$ & $1.6$ & $-7.5$ & $-4.6$ & $1.3$ & $-8.2$ & $-4.9$ & $1.3$ & $-8.2$ & $-8.6$ & $1.8$ & $-7.9$ & $-6.6$ & $1.4$ & $-8.6$ & $-7.5$ & $1.6$ & $-8.5$ \\
BC+Pantheon+ & $-14.8$ & $3.1$ & $-2.8$ & $-13.2$ & $2.9$ & $-3.2$ & $-13.6$ & $2.9$ & $-3.1$ & $-16.2$ & $3.0$ & $-4.6$ & $-13.4$ & $2.6$ & $-5.0$ & $-15.1$ & $2.8$ & $-4.7$ \\
BC+Pantheon++$H_0$ & $-8.0$ & $2.0$ & $-6.2$ & $-6.1$ & $1.6$ & $-6.7$ & $-6.6$ & $1.7$ & $-6.9$ & $-10.6$ & $2.2$ & $-6.9$ & $-7.7$ & $1.6$ & $-7.6$ & $-9.5$ & $2.0$ & $-8.2$ \\
BC+Union3 & $-14.1$ & $3.0$ & $-2.5$ & $-12.4$ & $2.7$ & $-2.2$ & $-13.0$ & $2.8$ & $-2.9$ & $-14.4$ & $2.7$ & $-4.5$ & $-12.9$ & $2.5$ & $-4.5$ & $-13.9$ & $2.7$ & $-3.8$ \\
BC+Union3+$H_0$ & $-4.4$ & $1.2$ & $-7.7$ & $-3.8$ & $1.1$ & $-7.9$ & $-3.4$ & $1.0$ & $-8.0$ & $-9.3$ & $1.9$ & $-7.8$ & $-11.8$ & $2.4$ & $-5.6$ & $-10.4$ & $2.1$ & $-6.5$ \\
\end{tabular}
\end{ruledtabular}
\end{table*}

Finally, we employ both Bayesian and frequentist statistics to identify the model preferred by the observational data. The values of Bayes factor and frequentist significance are summarized in Table~\ref{tab:lnB_withH0}. Additionally, we present the values of $\Delta \chi^2_{\rm min}$ for comparison in the same table. We find that all values of $\Delta \chi^2_{\rm min}$ are negative, with BC+SNe Ia yielding  $\Delta \chi^2_{\rm min} \lesssim -10$. When the SH0ES $H_0$ prior is added, the values of $\Delta \chi^2_{\rm min}$ increase significantly, though they remain less than zero. However, the Bayesian evidence consistently shows a  strong or very strong (definite only in a few cubic cases) preference for the  $\Lambda$CDM model over all  Lagrange polynomial reconstructions across all datasets. Specifically, all reconstructions are very strongly disfavored by BC+DES-Dovekie. Furthermore, BC+Pantheon+ provides very strong evidence only against the quartic-$y$ reconstruction, while BC+Union3 strongly disfavors the quartic reconstructions and definitely disfavors the cubic reconstructions. When the $H_0$ prior is applied, the log Bayes factors become more negative, further reinforcing the preference for the $\Lambda$CDM model over the reconstructions. Additionally, we find  that  the choice of interpolation variable ($z$, $y$, or $\zeta$) does not significantly influence the frequentist and Bayesian results. The frequentist significance, measured in terms of  $\sigma$, indicates that BC+SNe Ia favors the reconstruction at over a $2\sigma$ CL. However, imposing the SH0ES $H_0$ prior systematically reduces the statistical preference for the reconstructions. For example, in the BC+Union3 case, this prior reduces the preference for the cubic reconstructions from nearly $3\sigma$ to about $1\sigma$. This finding  is consistent with the results from the cubic reconstructed $\Delta q(z)$ curves, which show that once the $H_0$ prior is added, all best-fit $\Delta q(z)$ curves shift toward the $\Lambda$CDM prediction, indicating a loss of support for dynamical dark energy features. This suggests a tension between the SH0ES $H_0$ and the late-time dynamics of dark energy favored by the DESI DR2 BAO data~\cite{Pang2025SCPMA}. Overall, since increasing the interpolation order from $n=3$ to $n=4$ introduces additional freedom without a robust improvement in the model-comparison results, Occam's razor favors the cubic reconstruction as the simpler choice.
 
It is noteworthy that the Bayesian and frequentist model comparison results point in opposite directions in our analysis: the Bayesian evidence favors $\Lambda$CDM, whereas the frequentist method favors the Lagrange-interpolation reconstructions,  hinting at dynamical dark energy. Such a Bayesian-frequentist discrepancy is also reported by a recent fully Bayesian reanalysis of DESI DR2 BAO data~\cite{Ong2026arXiv}. They found that the combinations BAO+CamSpec (lensing) and BAO+CamSpec (lensing)+Pantheon+ yielded frequentist preferences for $w_0w_a$CDM, while the Bayesian evidence still favored $\Lambda$CDM. This discrepancy can be interpreted as an instance of the Jeffreys-Lindley paradox~\cite{Lindley1957Biome,Jeffreys1935PCPS,Robert2013arXiv,Wagenmakers2021arXiv}, which underscores a fundamental tension between Bayesian and frequentist approaches to hypothesis testing and reflects the Bayesian  penalty against extended models. Our results  align with this pattern: a moderate improvement in the best-fit $\chi^2$ does not necessarily indicate Bayesian support for the extended model.

\section{Conclusion and Discussion}\label{S5}

Recently, DESI Collaboration~\cite{Lodha2025PhRvD} conducted an extended analysis of dark energy constraints by combining BAO, CMB, and three compilations of SNe Ia. They found clear evidence that current data favor dynamical dark energy, particularly at low redshift. Their reconstructed $q(z)$ exhibits nontrivial features, including oscillatory behavior, and indicates a weaker cosmic acceleration at low redshift compared to the $\Lambda$CDM model. Motivated by these findings, we aim to probe possible dynamical features in the late-time cosmic expansion history by directly reconstructing the deviation of the deceleration parameter from the fiducial $\Lambda$CDM model using cubic and quartic Lagrange interpolation, based on the latest BAO, CMB, and DES-Dovekie/Pantheon+/Union3 datasets. To mitigate the impact of the nonuniform distribution of data in redshift, we carry out the reconstructions not only in $z$ but also in the $y$-redshift and the log-redshift. Additionally, we assess the impact of imposing the SH0ES $H_0$ prior~\cite{Riess2022ApJL} to investigate the possible tension between the dynamical behavior of dark energy favored by the DESI DR2 BAO data and the  $H_0$ measurements from SH0ES, as highlighted by Pang et al.~\cite{Pang2025SCPMA}. Finally, we utilize Bayesian evidence and frequentist significance to determine which Lagrange interpolation provides the best approximation.

For all reconstructions considered, the Hubble constant inferred from BC+SNe Ia is consistent with the Planck results while remaining in tension with the SH0ES results at $\geq4.3\sigma$. Regarding the present value of the deceleration parameter, $q_0$, Lagrange polynomial reconstructions based on BC and BC+SNe Ia indicate a weaker present-day acceleration than predicted by the $\Lambda$CDM model. The transition redshift $z_t$, together with the intermediate-redshift deceleration parameter $q_{\rm pivot}$ inferred from all reconstructions, suggests that cosmic acceleration began earlier than predicted by the $\Lambda$CDM model. Furthermore, all reconstructed $\Delta q(z)$ curves fitted to BC+SNe Ia show a clear low-redshift deviation from the $\Lambda$CDM prediction for both cubic and quartic reconstructions. The reconstructed $\Delta q(z)$ exhibits an overall oscillatory pattern around the $\Lambda$CDM model throughout the evolution; for cubic-$y$ and cubic-$\zeta$ reconstructions with the BC+Union3 data, cosmic acceleration has already reached its maximum at low $z$, and is beginning to slow down. Notably, the cubic-$y$ and cubic-$\zeta$ reconstructions constrained by BC+Union3 are broadly consistent with the findings reported by DESI Collaboration. However, once the SH0ES $H_0$ prior is imposed, the inferred $\Delta q(z)$ evolution in the cubic reconstructions shifts toward the $\Lambda$CDM model, significantly reducing the apparent low-$z$ discrepancy. Taken together, these results suggest that DESI BAO data point to dynamical behavior of dark energy, whereas the SH0ES prior drives the reconstruction toward the $\Lambda$CDM prediction, highlighting a tension between the SH0ES $H_0$ measurement and the DESI BAO data.

The reconstructed results presented in this paper indicate that the BC+SNe Ia dataset strongly supports a cosmic evolution that deviates from the $\Lambda$CDM model. However, this deviation is significantly reduced when the SH0ES measurement of $H_0$ is taken into account. The discrepancy between the Bayesian evidence in favor of $\Lambda$CDM and the frequentist significance supporting the Lagrange interpolation reconstructions suggests that current observations do not yet provide conclusive evidence to distinguish an evolving cosmic expansion history from that described by $\Lambda$CDM. Furthermore, we found that the $y$ and $\zeta$ variables should be employed in future reconstructions of cosmic evolution. Notably, only the cubic Lagrange interpolations using $y$ and $\zeta$ as variables, based on the BC+Union3 data, provide evidence that cosmic acceleration is slowing down. Therefore, our understanding of cosmic expansion at low redshifts requires further investigation through more precise future observations.

\begin{acknowledgments}
We thank Yang Liu for helpful discussions. We are grateful to the anonymous referees for their constructive comments and suggestions, which helped improve
the manuscript. This work was supported by the National Natural Science Foundation of China under Grants No. 12305056, No. 12275080, No. 12635002, No. 12075084, and No. 11505004, the Innovative Research Group of Hunan Province under Grant No. 2024JJ1006, the Cultivation Project for Young and Middle-aged Teachers in Provincial Colleges and Universities under Grant No. YQZD2024034, the Anhui Science and Technology University's Key Discipline Construction Fund~(XK-XJGY002), and the Natural Science Foundation of Anhui Province under Grant No. 1508085QA17.
\end{acknowledgments}

\bibliography{ref}

\end{document}